\documentstyle[epsf,graphics,epsfig]{mn}

\def\hrp{\mbox{\boldmath $\hat{r'}$}}
\def\vlg{\mbox{\boldmath $v_{LG}$}}
\def\r{\mbox{\boldmath $r$}}
\def\rp{\mbox{\boldmath $r'$}}
\def\s{\mbox{\boldmath $s$}}
\def\sp{\mbox{\boldmath $s'$}}
\def\C{\mbox{\boldmath $C$}}
\def\D{\mbox{\boldmath $D$}}

\def\be{\begin{equation}}
\def\ee{\end{equation}}
\def\bm{\begin{displaymath}}
\def\em{\end{displaymath}}
\def\Real{{\rm Re\,}}
\def\Imag{{\rm Im\,}}
\def\llangle{\left\langle}
\def\rrangle{\right\rangle}
\def\hompc{\,h\,{\rm Mpc}^{-1}}
\def\mpcoh{\,h^{-1}\,{\rm Mpc}}
\def\kms{\,{\rm km\,s^{-1}}}

\font\japit = cmti10 at 11truept

\title[The 2dF Galaxy Redshift Survey: Spherical Harmonics analysis of
fluctuations in the final catalogue] 
{\vglue-3.0truecm
\centerline{\japit Accepted for publication in Monthly Notices of the R.A.S.}
\vglue 2.5truecm\noindent
The 2dF Galaxy Redshift Survey: Spherical Harmonics analysis of
fluctuations in the final catalogue}

\date{Accepted for publication in MNRAS}
\begin{document}

\author[W.J.~Percival et~al.]{
\parbox[t]{\textwidth}{
Will J.\ Percival$^1$,
Daniel Burkey$^1$,
Alan Heavens$^1$,
Andy Taylor$^1$,
Shaun Cole$^2$, 
John A.\ Peacock$^1$,
Carlton M.\ Baugh$^2$,
Joss Bland-Hawthorn$^3$,
Terry Bridges$^{3,4}$, 
Russell Cannon$^3$, 
Matthew Colless$^3$, 
Chris Collins$^5$, 
Warrick Couch$^6$, 
Gavin Dalton$^{7,8}$,
Roberto De Propris$^9$,
Simon P.\ Driver$^9$, 
George Efstathiou$^{10}$, 
Richard S.\ Ellis$^{11}$, 
Carlos S.\ Frenk$^2$, 
Karl Glazebrook$^{12}$, 
Carole Jackson$^{13}$,
Ofer Lahav$^{10,14}$, 
Ian Lewis$^7$, 
Stuart Lumsden$^{15}$, 
Steve Maddox$^{16}$,
Peder Norberg$^{17}$,
Bruce A.\ Peterson$^9$, 
Will Sutherland$^{10}$,
Keith Taylor$^{11}$ (The 2dFGRS Team)}
\vspace*{6pt} \\ 
$^1$   Institute for Astronomy, University of Edinburgh, Royal Observatory, 
       Blackford Hill, Edinburgh EH9 3HJ, UK \\
$^2$   Department of Physics, University of Durham, South Road, Durham DH1 3LE, UK \\
$^3$   Anglo-Australian Observatory, P.O.\ Box 296, Epping, NSW 2121,
       Australia\\  
$^4$   Physics Department, Queen's University, Kingston, ON, K7L 3N6, Canada \\
$^5$   Astrophysics Research Institute, Liverpool John Moores University,  
       Twelve Quays House, Birkenhead L14 1LD, UK \\
$^6$   Department of Astrophysics, University of New South Wales, Sydney, 
       NSW 2052, Australia \\
$^7$   Department of Physics, University of Oxford, Keble Road, Oxford OX1 3RH, UK \\
$^8$   Space Science and Technology Division, Rutherford Appleton
       Laboratory, Chilton, Didcot OX11 0QX, UK \\
$^9$   Research School of Astronomy \& Astrophysics, The Australian 
       National University, Weston Creek, ACT 2611, Australia \\
$^{10}$Institute of Astronomy, University of Cambridge, Madingley Road,
       Cambridge CB3 0HA, UK \\
$^{11}$Department of Astronomy, Caltech, Pasadena, CA 91125, USA \\
$^{12}$Department of Physics \& Astronomy, Johns Hopkins University,
       Baltimore, MD 21218-2686, USA \\
$^{13}$CSIRO Australia Telescope National Facility, PO Box 76, Epping,
       NSW 1710, Australia\\
$^{14}$Department of Physics and Astronomy, University College London,
       Gower Street, London WC1E 6BT, UK\\
$^{15}$Department of Physics, University of Leeds, Woodhouse Lane,
       Leeds LS2 9JT, UK \\
$^{16}$School of Physics \& Astronomy, University of Nottingham,
       Nottingham NG7 2RD, UK \\
$^{17}$ETHZ Institut f\"ur Astronomie, HPF G3.1, ETH H\"onggerberg,
       CH-8093 Z\"urich, Switzerland \\
}

\date{}

\maketitle

\begin{abstract}
We present the result of a decomposition of the 2dFGRS galaxy
overdensity field into an orthonormal basis of spherical harmonics and
spherical Bessel functions. Galaxies are expected to directly follow
the bulk motion of the density field on large scales, so the absolute
amplitude of the observed large-scale redshift-space distortions
caused by this motion is expected to be independent of galaxy
properties. By splitting the overdensity field into radial and angular
components, we linearly model the observed distortion and obtain the
cosmological constraint $\Omega_m^{0.6}\sigma_8=0.46\pm0.06$. The
amplitude of the linear redshift-space distortions relative to the
galaxy overdensity field is dependent on galaxy properties and, for
$L_*$ galaxies at redshift $z=0$, we measure
$\beta(L_*,0)=0.58\pm0.08$, and the amplitude of the overdensity
fluctuations $b(L_*,0)\sigma_8=0.79\pm0.03$, marginalising over the
power spectrum shape parameters. Assuming a fixed power spectrum shape
consistent with the full Fourier analysis produces very similar
parameter constraints.
\end{abstract}


\section{introduction}

Analysis of galaxy redshift surveys provides a statistical measure of
the surviving primordial density perturbations. These fluctuations
have a well known dependency on cosmological parameters
(e.g. Eisenstein \& Hu 1998), and can therefore be used to constrain
cosmological models. The use of large scale structure as a
cosmological probe has acquired an increased importance in the new era
of high precision cosmology, which follows high-quality measurements
of the cosmic microwave background (CMB) power spectrum (Bennett
et~al. 2003; Hinshaw et~al. 2003). The extra information from galaxy
surveys helps to lift many of the degeneracies intrinsic to the CMB
data and enhances the scientific potential of both data sets
(e.g. Efstathiou et~al. 2002; Percival et~al. 2002; Spergel
et~al. 2003; Verde et~al. 2003).

In this paper we decompose the large-scale structure density
fluctuations observed in the 2dF Galaxy Redshift Survey (2dFGRS;
Colless et~al. 2001;2003) into an orthonormal basis of spherical
harmonics and spherical Bessel functions. In Percival et~al. (2001;
P01) we decomposed the partially complete 2dFGRS into Fourier modes
using the method outlined by Feldman, Kaiser \& Peacock (1994). In a
companion paper (Cole et~al. 2004; C04) we analyse the final catalogue
using Fourier modes. In P01 and C04, the Fourier modes were
spherically averaged and fitted with model power spectra convolved
with the spherically averaged survey window function. Redshift-space
distortions destroy the spherical symmetry of the convolved power and
potentially distort the recovered power from that expected with a
simple spherical convolution. Analysis of mock catalogues presented in
P01 and a detailed study presented in C04 showed that, in spite of
these complications, cosmological parameter constraints can still be
recovered from a basic Fourier analysis.

However, a decomposition into spherical harmonics and spherical Bessel
functions rather than Fourier modes distinguishes radial and angular
modes, and enables redshift-space distortions to be easily introduced
into the analysis method (without the far-field approximation, Kaiser
1987), as well as allowing for the effects of the radial selection
function and angular sky coverage (Heavens \& Taylor 1995; HT). The
down-side is that Spherical Harmonics methods are, in general, more
complex than Fourier techniques and are computationally more
expensive. This is particularly apparent when only a relatively small
fraction of the sky is to be modelled, as the observed modes are then
the result of a convolution of the true modes with a wide window
function. For nearly all sky surveys (e.g. IRAS surveys), correlations
between modes are reduced, and the window is narrower leading to a
reduced computational budget.

Consequently, a number of Spherical Harmonics decompositions have been
previously performed for the IRAS surveys. The primary focus of much
of the earlier work was the measurement of
$\beta(L,z)\equiv\Omega_m(z)^{0.6}/b(L,z)$, a measure of the increased
fluctuation amplitude caused by the linear movement of matter onto
density peaks and out from voids (Kaiser 1987). Here $\Omega_m(z)$ is
the matter density and $b(L,z)$ is a simplified measure of the
relevant galaxy bias. See Berlind, Narayanan \& Weinberg (2001) for a
detailed study of $\beta(L,z)$ measurements assuming more realistic
galaxy bias models.  

For the IRAS 1.2-Jy survey, HT and Ballinger, Heavens \& Taylor (1995)
found $\beta\sim1\pm0.5$ for fixed and varying power spectrum shape
respectively, and similar constraints were also found by Fisher,
Scharf \& Lahav (1994), Fisher et~al. (1995). However, Cole, Fisher \&
Weinberg (1995) found $\beta=0.52\pm0.13$ and Fisher \& Nusser (1996)
found $\beta=0.6\pm0.2$ for the 1.2-Jy survey using the
quadrupole-to-monopole ratio for the decomposition of the power
spectrum into Legendre polynomials. No explanation for the apparent
discrepancy between these results has yet been found, although we note
that the results are consistent at approximately the 1-$\sigma$ level
if the large errors are taken into account for the Spherical Harmonics
decompositions.

The IRAS Point Source Catalogue Redshift Survey (PSCz; Saunders et~al.
2000) has also been analysed using a Spherical Harmonics decomposition
by a number of authors (Tadros et~al. 1999; Hamilton, Tegmark \&
Padmanabhan 2000; Taylor et~al. 2001) who found $\beta\sim0.4$. More
recently, Tegmark, Hamilton \& Xu (2002) presented an analysis using
spherical harmonics to decompose the first 100k redshifts released
from the 2dFGRS and found $\beta=0.49\pm0.16$ for the
$b_{\scriptscriptstyle\rm J}$ selected galaxies in this survey,
consistent with the $\xi(\sigma,\pi)$ analyses of Peacock
et~al. (2001) and Hawkins et~al. (2003). The measured $\beta$
constraints are expected to vary between samples through the
dependence on the varying galaxy bias. For example, by analysing the
bispectrum of the PSCz survey Feldman et~al. (2001) found a smaller
large-scale bias than a similar analysis of the 2dFGRS by Verde
et~al. (2002).

In addition to the linear distortions, random galaxy motions within
galaxy groups produce the well known Fingers-of-God effect where
structures are elongated along the line-of-sight. These random
motions mean that the observed power is a convolution of the
underlying power with a narrow window. The observed power therefore
depends on the form of this window and the amplitude of the velocity
dispersion as a function of scale.

The 2dFGRS and Sloan Digital Sky Survey (SDSS; Abazajian et~al. 2004)
cover sufficient volume that it is now possible to recover information
about the shape of the power spectrum in addition to the
redshift-space distortions (P01; C04; Tegmark, Hamilton \& Xu 2002;
Tegmark et~al. 2003a). However, the decreased random errors (cosmic
variance) of these new measurements means that systematic
uncertainties have become increasingly important. In particular,
galaxies are biased tracers of the matter distribution: the relation
between the galaxy and mass density fields is probably both nonlinear
and stochastic to some extent (e.g. Dekel \& Lahav 1999), so that the
power spectra of galaxies and mass differ in general. Assuming that
the bias tends towards a constant on large scales, then we can write
$P_g(k)=b^2P_m(k)$, where subscripts $m$ and $g$ denote matter and
galaxies respectively. For the 2dFGRS galaxies, although the average
bias is close to unity (Lahav et~al. 2002; Verde et~al. 2002), the
bias is dependent on galaxy luminosity (Norberg et~al. 2001; 2002a;
Zehavi et~al. 2002 find a very similar dependence for SDSS galaxies),
with $\llangle b(L,z)/b(L_*,z)\rrangle = 0.85 + 0.15 L/L_*$ where the
bias $b(L,z)$ is assumed to be a simple function of galaxy luminosity
and $L_*$ is defined such that $M_{b_{\rm J}}-5\log_{10}h=-19.7$
(Norberg et~al. 2002b). Because average galaxy luminosity is a function
of distance, this bias can distort the shape of the recovered power
spectrum (Tegmark et~al. 2003a; Percival, Verde \& Peacock 2004).

In this paper we decompose the final 2dFGRS catalogue into an
orthonormal basis of spherical harmonics and spherical Bessel
functions and fit cosmological models to the resulting mode
amplitudes. To compress the modes we adopt a modified
Karhunen-Lo\`{e}ve (KL) data compression method that separates angular
and radial modes (Vogeley \& Szalay 1996; Tegmark, Taylor \& Heavens
1997; Hamilton, Tegmark \& Padmanabhan 2000; Tegmark, Hamilton \& Xu
2002). We also include a consistent correction for
luminosity-dependent bias that includes the effect of this bias on
both the measured power and fitted models. We have performed two fits
to the recovered modes. First we measured the galaxy power spectrum
amplitude, $b(L_*,0)\sigma_8$ and the linear infall amplitude,
$\beta(L_*,0)$ for a fixed power spectrum shape. We then considered
fitting a more general selection of cosmological models to these data.

A detailed analysis of the internal consistency of the 2dFGRS
catalogue with respect to measuring $P(k)$ was performed using a
Fourier decomposition of the galaxy density field and is presented in
C04. This analysis included looking at the effect of changing the
calibration, maximum redshift, weighting, region, and galaxy colour
range considered. This work is not duplicated using our decomposition
technique, and we instead refer the interested reader to that
paper. Tests presented in this paper are primarily focused on the
analysis method, although we consider the effect of the catalogue
calibration in Section~\ref{sec:tests}.

The layout of this paper is as follows. In Section~\ref{sec:2dFcat} we
describe the 2dFGRS catalogue analysed, and in
Section~\ref{sec:mockcat} we consider mock catalogues used to test our
analysis method. A brief overview of the methodology is presented in
Section~\ref{sec:method_overview}. A full description of the Spherical
Harmonics method used is provided in Appendix~\ref{app:A}. The results
are presented for both the 2dFGRS and mock catalogues in
Sections~\ref{sec:results_fixshape} \&~\ref{sec:results_varshape}. A
discussion of various tests performed is given in
Section~\ref{sec:tests}. We conclude in Section~\ref{sec:discussion}.

\section{the 2\lowercase{d}FGRS catalogue}  \label{sec:2dFcat}

In this work, we consider the final 2dFGRS release catalogue. However,
the formalism adopted is simplified if we consider a catalogue with a
selection function that is separable in radial and angular directions
(see Appendix~\ref{app:A} for details of the formalism). There are two
complications in the 2dFGRS catalogue that cause departures from such
behaviour (as discussed in Colless et~al. 2001;2003).

\begin{enumerate}

  \item The photometric calibration of the UKST plates from which the
  2dFGRS sample was drawn and the extinction correction have been
  revised after the initial sample selection. Because revision of the
  galaxy magnitudes and the angular magnitude limit are required, this
  means that the survey depth varies across the sky.

  \item Due to seeing variations between observations, the overall
  completeness varies with apparent magnitude with a form that depends
  on the field redshift completeness. This is characterized by a
  parameter $\mu$, with the varying completeness given by
  $c_z(m,\mu)=0.99[1-\exp(m-\mu)]$ (Colless et~al. 2001).

\end{enumerate}

Rather than adapt the formalism, we have chosen to use a reduced
version of the 2dFGRS release catalogue with a window function that is
separable in radial and angular directions. These issues were also
discussed with reference the 100k release catalogue by Tegmark,
Hamilton \& Xu (2002) whose method also required a sample with window
function and weights separable in radial and angular
directions. Correcting for these effects is relatively
straightforward, if a little painful as we have to remove galaxies
with valid redshifts from the analysis. First we need to select a
uniform revised magnitude limit at which to cut the
catalogue. Galaxies fainter than this limit are removed from the
revised catalogue, as are galaxies that were selected using an actual
magnitude limit that was brighter than the revised limit. Selecting
the revised magnitude limit at which to cut the catalogue is a
compromise between covering as large an angular region as possible
(resulting in a narrow angular window function), covering as large a
weighted volume as possible (reducing cosmic variance), or retaining
as many galaxies as possible (reducing shot noise). However, we can
model variations in the angular window function and, in Percival
et~al. (2001), we showed that the 2dFGRS sample is primarily cosmic
variance limited. We therefore chose the magnitude limit to maximize
the effective volume of the survey.

The random fields, a number of circular 2-degree fields randomly
placed in the low extinction regions of the southern APM galaxy survey
were excluded from our analysis, in order to focus on two contiguous
regions with well-behaved selection functions. These two regions of
the survey, one near the north galactic pole (NGP) and another near
the south galactic pole (SGP) were analysed separately, and
optimization of the magnitude limit was performed for each region
independently. The resulting limits are given in
Table~\ref{tab:cats}. In order to correct for the magnitude dependent
completeness, we removed a randomly selected sample of the bright
galaxies in order to provide uniform completeness as a function of
magnitude.

\begin{table}
  \caption{Limiting extinction-corrected magnitudes, numbers of
  galaxies, and assumed radial selection-function parameters for each
  of the two 2dFGRS regions modelled. The parameters controlling the
  radial distribution are defined by Eq.~\ref{eq:fz}.
  \label{tab:cats}}

  \centering \begin{tabular}{cccccc} \hline
  region & $M_{\rm lim}$ & N$_{\rm gal}$ & $z_c$ & $b$  & $g$  \\
  \hline
  SGP    & 19.29 & 84824 & 0.130 & 2.21 & 1.34 \\
  NGP    & 19.17 & 57932 & 0.128 & 2.45 & 1.24 \\
  \hline
  \end{tabular}
\end{table}

\begin{figure}
  \setlength{\epsfxsize}{0.7\columnwidth} 
    \centerline{\epsfbox{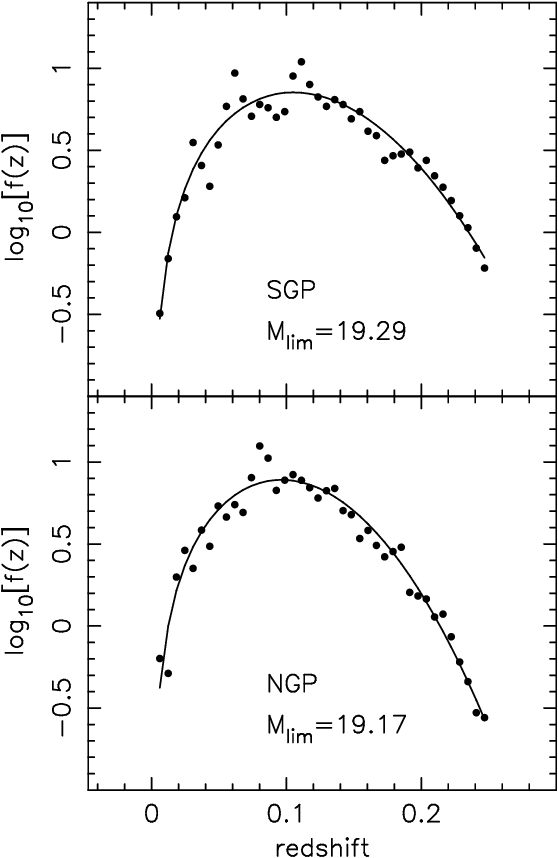}}

  \caption{Redshift distribution of the reduced galaxy catalogues for
  the two regions considered (solid circles), compared with the best
  fit redshift distribution for each of the form given by
  Eq.~\ref{eq:fz}. The magnitude limit adopted for each sample is
  given in each panel.}

  \label{fig:fz}
\end{figure}

The redshift distribution of each sample was matched using a
function of the form
\begin{equation}
  f(z)\propto z^g \exp\left[-\left(\frac{z}{z_c}\right)^b\right],
    \label{eq:fz}
\end{equation}
where the parameters $z_c$, $b$ \& $g$ were calculated by fitting to
the weighted (Eq.~\ref{eq:wght}) redshift distribution, calculated in
$40$ bins equally spaced in $z$. These resulting parameter values are
given in Table~\ref{tab:cats}, and the redshift distributions are
compared with the fits in Fig.~\ref{fig:fz}. In addition to the radial
and angular distributions of the sample, we also need to match the
normalization of the catalogue to the expected distribution. We choose
to normalize each catalogue by matching $\int d\r \bar{n}(\r)w(\r)$,
where $\bar{n}(\r)$ is the expected galaxy distribution function and
$w(\r)$ is the weight applied to each galaxy (Eq.~\ref{eq:wght}), for
reasons described in Percival, Verde \& Peacock (2004).

\section{the mock catalogues}  \label{sec:mockcat}

As a test of the Spherical Harmonics procedure adopted, we have
applied our method to recover parameters from the 22 LCDM03 Hubble
Volume mock catalogues available from {\tt
http://star-www.dur.ac.uk/{\tt\char'176}cole/mocks/main.html} (Cole
et~al. 1998). These catalogues were calculated using an
empirically-motivated biasing scheme to place galaxies within $N$-body
simulations, and were designed to cover the 2dFGRS volume. We have
applied the same magnitude and completeness cuts to these data, as
applied to the 2dFGRS catalogue (Section~\ref{sec:2dFcat}). In order
to allow for slight variations between the redshift distribution of
the mocks and the 2dFGRS catalogue, we fit the redshift distribution
of the mock catalogues independently from the 2dFGRS data. Because we
adopt the magnitude limits used for the 2dFGRS data, the NGP and SGP
regions in the mock catalogues have different redshift distributions
and these are fitted separately. For simplicity, we assume a single
expected redshift distribution for each region for all of the mocks,
calculated by fitting to the redshift distribution of the combination
of all of the mocks. The number of galaxies in each catalogue is
sufficient that the model of $f(z)$ only changes slightly when
considering either catalogues individually, or the combination of all
22 catalogues. This change is sufficiently small that it does not
significantly alter either the recovered parameters from the mock
catalogues or their distribution.

We use these mock catalogues in a number of ways. By comparing the
average recovered parameters and known input parameters of the
simulations, we test for systematic problems with the method. In fact,
we did not analyse the 2dFGRS data until we had confirmed the validity
of the method through application to these mock catalogues. We test
our recovery of the linear redshift-space distortion parameter
$\beta(L_*,z)$ by analysing mocks within which galaxy peculiar
velocities were altered (Section~\ref{sec:Btest}). Additionally, we
use the distribution of recovered values to test the confidence
intervals that we can place on recovered parameters
(Section~\ref{sec:confidence}).

\section{method overview}  \label{sec:method_overview}

The use of Spherical Harmonics to decompose galaxy surveys dates back
to Peebles (1973), and is a powerful technique for statistically
analysing the distribution of galaxies. The formalism used in this
paper is based in part on that developed by HT and described by Tadros
et~al. (1999). However, there are some key differences and extensions,
which warrant the full description given in Appendix~\ref{app:A}. In
this section we outline the procedure for a non-specialist reader.

The galaxy density field was decomposed into an orthonormal basis
consisting of spherical Bessel functions and spherical harmonics. In
general, we refer to this as a Spherical Harmonics decomposition. As in
P01 \& C04, we decomposed the density field in terms of proper
distance and therefore needed to assign a radial distance to each
galaxy. For this, we adopted a flat cosmology with $\Omega_m=0.3$ and
$\Omega_\Lambda=0.7$. The dependence of the recovered power spectrum
and $\beta(L_*,z)$ on this ``prior'' is weak, and was explored in
P01. We assume a constant galaxy clustering (CGC) model, where the
amplitude of galaxy clustering is independent of redshift, although it
is dependent on galaxy luminosity through the relation of Norberg
et~al. (2001) given in Eq.~\ref{eq:bobstar}. This relates the
clustering amplitude of galaxies of luminosity $L$ to that of $L_*$
galaxies, and by weighting each galaxy by the reciprocal of this
relation, we correct for luminosity-dependent bias.

The Spherical Harmonics decomposition of the mean expected
distribution of galaxies is then subtracted, calculated using a fit to
the radial distribution and an angular mask (this was modelled using a
random catalogue in the Fourier analyses of P01 \& C04). This converts
from a decomposition of the density field to the overdensity field.

In the Fourier based analyses of P01 and C04, we modelled the observed
power spectrum. In the analysis presented in this paper we instead
model the transformed overdensity field. The expected value of the
transform of the overdensity field for any cosmological model is zero
by definition. Consequently, apart from a weak dependence on a prior
cosmological model hard-wired into the analysis method, the primary
dependence on cosmological parameters is encapsulated in the
covariance matrix used to determine the likelihood of each model.

The primary difficulty in calculating the covariance matrix for a
given cosmological model is correctly accounting for the geometry of
the 2dFGRS sample. This results in a significant convolution of the
true power, and is performed as a discrete sum over Spherical Harmonic
modes in a computationally intensive part of the adopted procedure. To
first order, the large-scale redshift-space distortions are linearly
dependent on the density field, and we can therefore split the
covariance matrix into four components corresponding to the mass-mass,
mass-velocity and velocity-velocity power spectra (cf. Tegmark,
Hamilton \& Xu 2002; Tegmark et~al. 2003a) and the shot noise. This is
discussed after Eq.~\ref{eq:Dsplit2} in Appendix~\ref{app:A}. The
velocity component of the covariance matrix is dependent on matter
density field rather than the galaxy density field, and we include a
correction for the linear evolution of this field. For this we assume
that $\Omega_m=0.3$ and $\Omega_\Lambda=0.7$ although the resulting
covariance matrix is only weakly dependent on this ``prior''.

We include the contribution from the small-scale velocity dispersion
of galaxies by undertaking an additional convolution of the radial
component of these matrices. We choose to model scales where the
small-scale velocity dispersion does not contribute significantly to
the overdensity field, and demonstrate this weak dependence in
Section~\ref{sec:ssvd}.

The transformation between Fourier modes and Spherical Harmonics is
unitary, so each Spherical Harmonic mode corresponds directly to a
particular Fourier wavelength.  Within our chosen decomposition of the
density field, there are $86667$ Spherical Harmonic modes with
$0.02<k<0.15\hompc$, and it is impractical to use all of these modes
in a likelihood analysis as the inversion of an $86667\times86667$
matrix is slow and may be unstable for a problem such as this. The
modes were therefore compressed, leaving $1223$ \& $1785$ combinations
of modes for the NGP and SGP respectively. The data compression
procedure adopted was designed to remove nearly degenerate modes,
which could cause numerical problems, and to optimally reduce the
remaining data. The compressed data, and the corresponding covariance
matrices are then combined to calculate the likelihood of a given
model assuming Gaussian statistics.

Only $\sim 5\,\%$ of the computer code used in the PSCz analysis of
Tadros et~al. (1999) was reused in the current work, as both revision
of the method and a significant speed-up of the process were required
to model the 2dFGRS. In particular, the geometry of the 2dFGRS sample
means that the convolution to correct for the survey window function
requires calculation for a larger number of modes than all-sky surveys
such as the PSCz, and the method consequently takes longer to
run. Because of this revision, the method required thorough testing,
both by analysing mock catalogues and by considering the specific
tests described in Section~\ref{sec:tests}.

\section{results}  \label{sec:results}

\begin{figure*}
  \setlength{\epsfxsize}{0.7\textwidth} 
    \centerline{\epsfbox{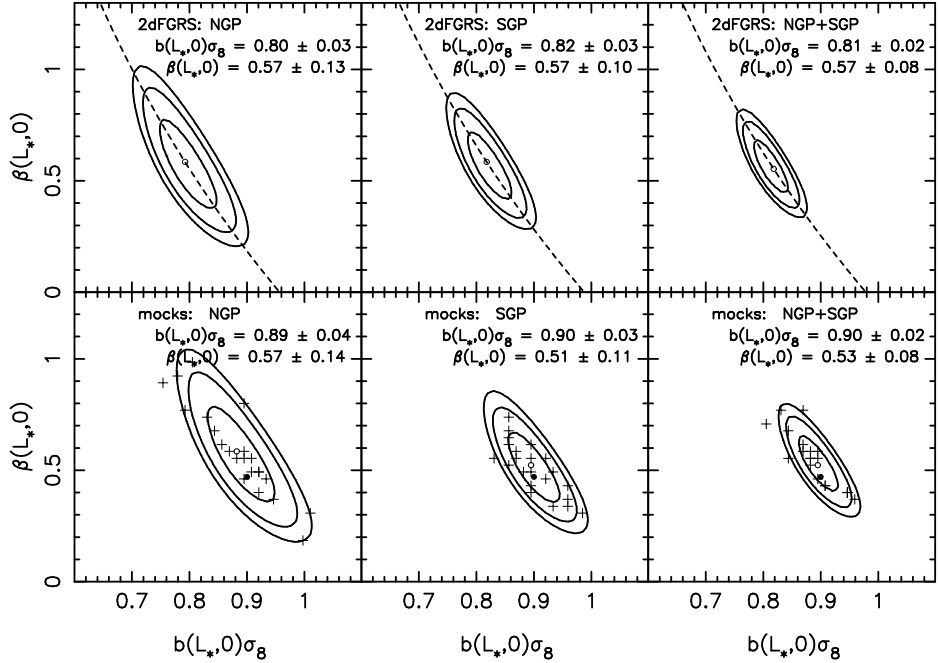}}

  \caption{Likelihood contours for the recovered $b(L_*,0)\sigma_8$
  and $\beta(L_*,0)$ assuming a fixed $\Lambda$CDM power spectrum
  shape. Solid lines in the top row show the recovered contours from
  the 2dFGRS, while the bottom row gives the average recovered
  contours from the $\Lambda$CDM mock catalogues. Contours correspond
  to changes in the likelihood from the maximum of $2\Delta\ln{\cal
  L}=2.3, 6.0, 9.2$. These values correspond to the usual
  two-parameter confidence of 68, 95 and 99 per cent. The open circle
  marks the ML position, while the solid circle marks the true
  parameters for the mock catalogues. The crosses give the ML
  positions for the 22 mock catalogues. Note that on average $57\,\%$
  of the crosses lie within the $2\Delta\ln{\cal L}=2.3$ contour for
  the NGP and SGP mock catalogues. The chosen modes are not
  independent, although they are orthogonal, so we cannot assume that
  $\ln{\cal L}$ has a $\chi^2$ distribution. See
  Section~\ref{sec:confidence} for a further discussion of the
  confidence intervals that we place on recovered parameters. The
  dashed lines plotted in the upper panels give the locus of models
  with constant redshift space power spectrum amplitude (see text for
  details). \label{fig:sig8_beta}}
\end{figure*}

\begin{figure*}
  \setlength{\epsfxsize}{0.7\textwidth} 
    \centerline{\epsfbox{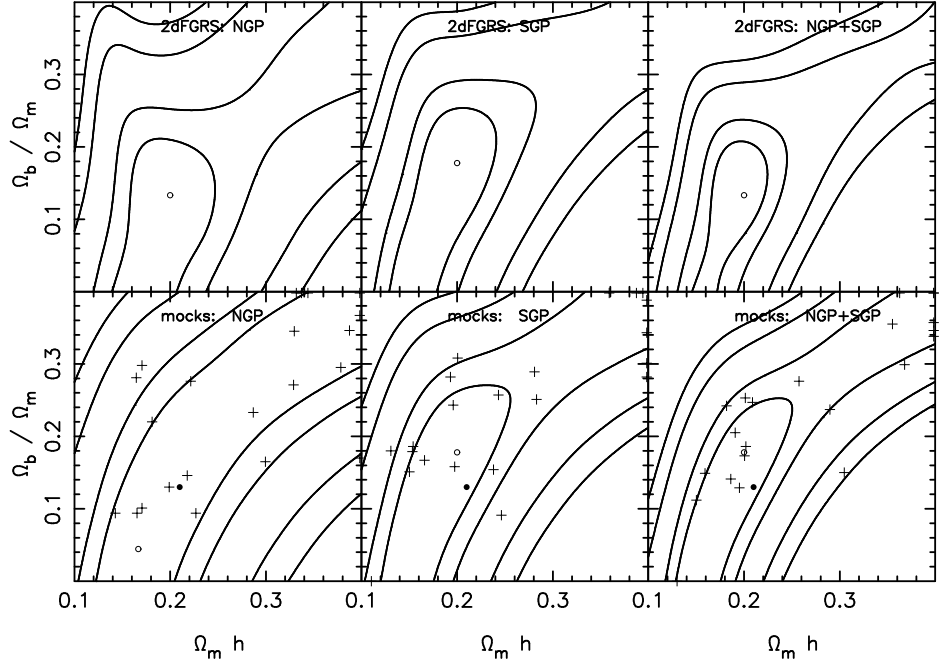}}

  \caption{Likelihood contours for $\Omega_mh$ and $\Omega_b/\Omega_m$
  assuming a $\Lambda$CDM power spectrum with $h=0.72$ and
  $n_s=1.0$. We have marginalised over the power spectrum amplitude
  and $\beta(L_*,0)$. Solid lines in the top row show the recovered
  contours from the 2dFGRS, while the bottom row gives the average
  recovered contours from the $\Lambda$CDM mock catalogues. Contours
  correspond to changes in the likelihood from the maximum of
  $2\Delta\ln{\cal L}=1.0, 2.3, 6.0, 9.2$. In addition to the
  contours plotted in Fig.~\ref{fig:sig8_beta}, we also show the
  standard one-parameter 68 per cent confidence region to match with
  figure~5 in P01. The open circle marks the ML position. As in P01,
  we find a broad degeneracy in the ($\Omega_mh$, $\Omega_b/\Omega_m$)
  plane, which is weakly lifted with a low baryon fraction favoured
  for the 2dFGRS data. These parameter constraints are less accurate
  than those derived in C04 as we use less data, and we limit the
  number of modes used. ML positions for the 22 mock catalogues are
  shown by the crosses. It can be seen that a number of the mock
  catalogues have likelihood surfaces that are not closed, with the ML
  position being at one edge of the parameter space
  considered. However, these mocks all follow the general degeneracy
  between models with the same $P(k)$ shape. \label{fig:omh_bf}}
\end{figure*}

\begin{figure*}
  \setlength{\epsfxsize}{0.7\textwidth} 
    \centerline{\epsfbox{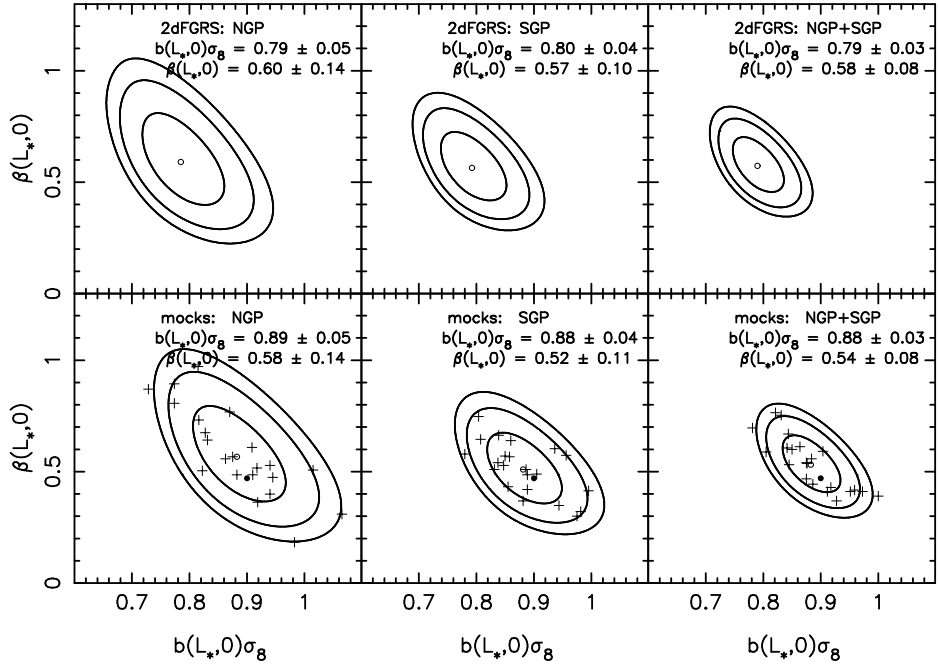}}

  \caption{As Fig.~\ref{fig:sig8_beta}, but now marginalising over the
  power spectrum shape as parameterized by $\Omega_mh$ and
  $\Omega_b/\Omega_m$. As can be seen, allowing for different power
  spectrum shapes only increases the errors on $b(L_*,0)\sigma_8$ and
  $\beta(L_*,0)$ slightly. The relative interdependence between the
  power spectrum shape and $b(L_*,0)\sigma_8$ and $\beta(L_*,0)$ is
  considered in more detail in Fig.~\ref{fig:param_degen}.
  \label{fig:sig8_beta_marg}}
\end{figure*}

Results are presented for the 2dFGRS catalogue described in
Section~\ref{sec:2dFcat}, and for the mock catalogues described in
Section~\ref{sec:mockcat}. Parameter constraints were derived fitting
to modes with $0.02<k<0.15\hompc$, the range considered in
P01. Because the Spherical Harmonics method includes the effects of
the small-scale velocity dispersion and uses a non-linear power
spectrum, we could in principle extend the fitted $k$-range to smaller
scales. However, our derivation of the covariance matrix is only based
on cosmic variance and shot noise. No allowance is made for systematic
offsets caused by our modelling of small-scale effects (velocity
dispersion, non-linear power and a possible scale-dependent galaxy
bias). Consequently, it is better to avoid regions in $k$-space that
are significantly affected by these complications, rather than assume
that we can model these effects perfectly. Additionally, the number of
modes that can be analysed is limited by computation time and the
large-scale $k$-range selected includes most of cosmological signal
and follows Gaussian statistics.

The Spherical Harmonics method involves a convolution of the window
and the model power over a large number of modes
(Eq.~\ref{eq:D1}). For a fixed power spectrum shape, the covariance
matrix can be written as a linear sum of four components with
different dependence on $b(L_*,0)\sigma_8$ and $\beta(L_*,0)$. It is
straightforward to store these components and these parameters can be
fitted without having to perform the convolution for each set of
parameters. In Section~\ref{sec:results_fixshape} we consider a fixed
power spectrum shape, and present results fitting $b(L_*,0)\sigma_8$
\& $\beta(L_*,0)$ to the 2dFGRS and mock catalogue data.

In an analysis of the power spectrum shape, separate convolutions are
required for each model $P(k)$. This would be computationally very
expensive, but can be circumvented by discretising the model $P(k)$ in
$k$ and performing a single convolution for each $k$-component. In
Section~\ref{sec:results_varshape} we fit to the power spectrum shape,
assuming a step-wise $P(k)$ in this way. First, we describe the set of
models to be considered.

\subsection{Cosmological parameters}  \label{sec:parameters}

A simple model is assumed for galaxy bias, with the galaxy overdensity
field assumed to be a multiple (the bias $b[L,0]$) of the present day mass
density field
\be
  \delta(L,\r)=b(L,0)\delta({\rm mass},\r),
\ee
at least for the survey smoothed near our upper wavenumber limit of
$0.15\hompc$. In the constant galaxy clustering model, the redshift
dependence of $b(L,z)$ is assumed to cancel that of the mass density
field so that $\delta(L,\r)$ is independent of redshift. Although
galaxy bias has to be more complicated in detail, we may hope that
there is a ``linear response limit'' on large scales: those probed in
the analysis presented in this paper. In the stochastic biasing
framework proposed by Dekel \& Lahav (1999), the simple model
corresponds to a dimensionless galaxy-mass correlation coefficient
$r_g=1$. Wild et~al. (2004) show that the correlation between
$\delta(L,\r)$ from different types of galaxies have $r_g>0.95$.

Modelling the expansion of the density field in spherical harmonics
and spherical Bessel functions is dependent on the linear
redshift-space distortions parameterized by
$\beta(L_*,z)\simeq\Omega_m(z)^{0.6}/b(L_*,z)$, a function of galaxy
luminosity and epoch. The evolution of this parameter is dependent on
that of the matter density $\Omega_m(z)$ and the galaxy bias
$b(L_*,z)$. These effects are included in the method and are
considered in Sections~\ref{sec:bias_model}
\&~\ref{sec:method_main}. The recovered expansion is also dependent on
the velocity dispersion $\sigma_{\rm pair}$ and model assumed for the
Fingers-of-God effect (see Section~\ref{sec:ssvd}).

We parameterise the shape of the power spectrum of $L_*$ galaxies with
the Hubble constant $h$ in units of $100 \,{\rm
km\,s^{-1}\,Mpc^{-1}}$, the scalar spectral index $n_s$, and the
matter density $\Omega_m$ through $\Omega_mh$ and the fraction of
matter in baryons $\Omega_b/\Omega_m$. The contribution to the matter
budget from neutrinos is denoted $\Omega_\nu$. The matter power
spectrum is normalized using $\sigma_8$, the present day rms linear
density contrast averaged over spheres of $8\mpcoh$ radius.

The shape of the power spectrum to current precision is only weakly
dependent on $h$, and only sets a strong constraint on a combination
of $\Omega_b/\Omega_m$, $\Omega_mh$, $\Omega_\nu$ and $n_s$. In this
paper, we only consider the very simple model of a constrained flat,
scale-invariant adiabatic cosmology with Hubble parameter $h=0.72$,
and no significant neutrino contribution $\Omega_\nu=0$. We show that
this model is consistent with our analysis, as it is with recent CMB
and LSS data sets (e.g. Spergel et~al. 2003; Tegmark et
al. 2003b). Additionally, we use $\Omega_b/\Omega_m$ \&~$\Omega_mh$ to
marginalise over the shape of the power spectrum when considering
$\beta(L_*,0)$ and $b(L_*,0)\sigma_8$, and marginalise over
$0<\Omega_b/\Omega_m<0.4$ and $0.1<\Omega_mh<0.4$. Given the precision
to which the shape of the power spectrum can be constrained, there is
an almost perfect degeneracy between $\Omega_b/\Omega_m$, $\Omega_mh$
and $n_s$. For $n_s\neq1$, to first order in $n_s$, our best-fit
values of $\Omega_b/\Omega_m$ and $\Omega_mh$ would change by
$0.46(n_s-1)$ and $0.34(1-n_s)$ respectively.

\subsection{Results for fixed power spectrum shape}  
  \label{sec:results_fixshape}

In this Section we fit $\beta(L_*,0)$ and $b(L_*,0)\sigma_8$ to the
data assuming a concordance model power spectrum with
$\Omega_mh=0.21$, $\Omega_{\rm b}/\Omega_{\rm m}=0.15$, $h=0.72$ \&
$n_{\rm s}=1$, consistent with the recent WMAP results (Spergel
et~al. 2003), and close to the true parameters of the Hubble volume
mocks. Likelihood contours for $b(L_*,0)\sigma_8$ and $\beta(L_*,0)$
are presented in Fig.~\ref{fig:sig8_beta} for the 2dFGRS and mock
catalogues. The primary degeneracy between these parameters arises
because $b(L_*,0)\sigma_8$ is a measure of the total power, combining
radial and angular modes. Increasing $\beta(L_*,0)$ beyond the
best-fit value increases the power of the model radial modes,
requiring a decrease in the overall power to approximately fit the
data. In order to show that this degeneracy corresponds to models with
the same redshift-space power spectrum amplitude, the dashed lines in
Fig.~\ref{fig:sig8_beta} show the locus of models with the same
redshift-space power spectrum amplitude as the maximum likelihood
solution. Here, the redshift-space and real-space power spectra,
represented by $P_s$ and $P_r$, are assumed to be related by
\be 
  P_s = (1+\frac{2}{3}\beta+\frac{1}{5}\beta^2)P_r.
    \label{eq:zspacepower} 
\ee
However, we still find tight constraints with
$b(L_*,0)\sigma_8=0.81\pm0.02$ and $\beta(L_*,0)=0.57\pm0.08$. In
fact, in Section~\ref{sec:results_varshape} we marginalise over a
range of model power spectra shapes and show that these constraints
are not significantly expanded when the shape of the power spectrum is
allowed to vary.

\subsection{Results without prior on the power spectrum shape}  
  \label{sec:results_varshape}

\begin{figure}
  \setlength{\epsfxsize}{0.9\columnwidth} 
    \centerline{\epsfbox{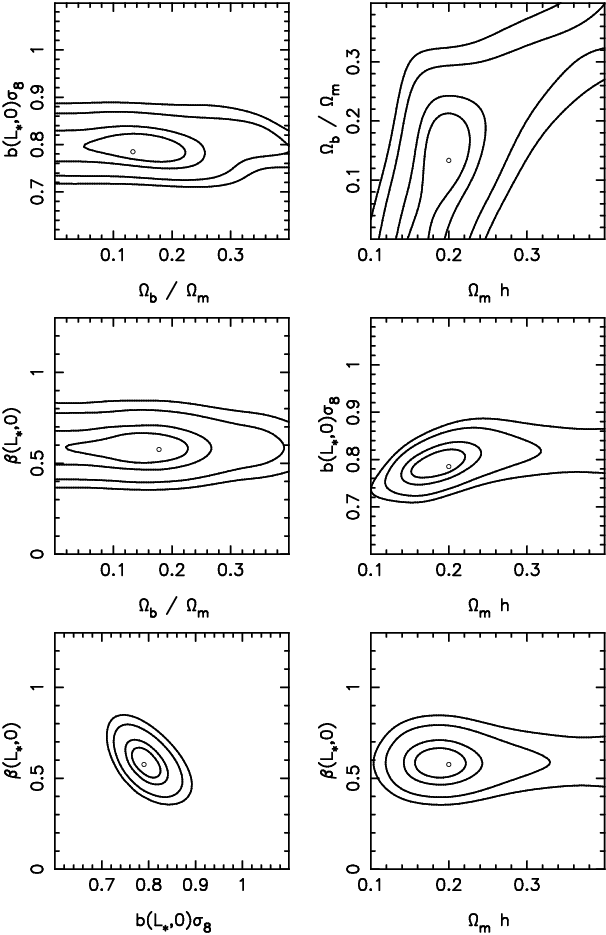}}

  \caption{Contour plots showing changes in the likelihood from the
  maximum of $2\Delta\ln{\cal L}=1.0, 2.3, 6.0, 9.2$ for different
  parameter combinations for the combined likelihood from the 2dFGRS
  NGP and SGP catalogues, assuming a $\Lambda$CDM power spectrum with
  $h=0.72$ and $n_s=1.0$. There are four parameters in total, and in
  each plot we marginalise over the two other parameters. The primary
  degeneracy arises between $\Omega_mh$ and $\Omega_b/\Omega_m$, and
  corresponds to similar power spectrum shapes. $b(L_*,0)\sigma_8$ is
  also degenerate with $\Omega_mh$, although $\beta(L_*,0)$ is
  independent of the power spectrum shape.  \label{fig:param_degen}}
\end{figure}

In Fig.~\ref{fig:omh_bf} we present likelihood contours for
$\Omega_mh$ and $\Omega_b/\Omega_m$ assuming a $\Lambda$CDM power
spectrum with fixed $n_s=1.0$, and marginalising over the power
spectrum amplitude $b(L_*,0)\sigma_8$ and $\beta(L_*,0)$. Apart from
the implicit dependence via $\Omega_mh$, there is virtually no
residual sensitivity to $h$, so we set it at the Hubble key project
value of $h=0.72$ (Freedman et~al. 2001). Contours are shown for the
recovered likelihood calculated using the NGP \& SGP catalogues and
from the combination of the two. We present the measured likelihood
surface from the 2dFGRS catalogue and the average likelihood surface
recovered from the mock catalogues. For both the 2dFGRS and the mocks,
there is a broad degeneracy between $\Omega_mh$ and
$\Omega_b/\Omega_m$, corresponding to models with similar power
spectrum shape. This degeneracy is partially lifted by the 2dFGRS
data, with a low baryon fraction favoured.
Fig.~\ref{fig:sig8_beta_marg} shows a similar plot for
$b(L_*,0)\sigma_8$ and $\beta(L_*,0)$, marginalising over the power
spectrum shape (parameterized by $\Omega_mh$ and
$\Omega_b/\Omega_m$). Although this increases the size of the allowed
region, the increase is relatively small, and we find
$\beta(L_*,0)=0.58\pm0.08$, and $b(L_*,0)\sigma_8=0.79\pm0.03$.

For the 2dFGRS catalogue, we present likelihood surfaces for all
parameter combinations in our simple 4 parameter model in
Fig.~\ref{fig:param_degen}. This plot shows that there is a degeneracy
between $b(L_*,0)\sigma_8$ and $\Omega_mh$ (as discussed for example
in Lahav et~al. 2002). However, $\beta(L_*,0)$ appears to be
independent of the power spectrum shape.

\subsection{Confidence intervals for parameters}  \label{sec:confidence}

Although the modes used are uncorrelated because of the
Karhunen-Lo\`{e}ve data compression (Section~\ref{sec:compression}),
they are not independent, and we cannot assume that $\ln{\cal L}$ has
a $\chi^2$ distribution. However, we can still choose to set fixed
contours in the likelihood as our confidence limits and simply need to
test the amplitude of the contours to be chosen. Luckily, we have 22
mock catalogues from which we can estimate confidence intervals. For
$b(L_*,0)\sigma_8$ and $\beta(L_*,0)$, with a fixed power spectrum,
$57\,\%$ of the data points lie within the $2\Delta\ln{\cal L}=2.3$
average contour for the NGP and SGP mock catalogues. Marginalising
over the power spectrum shape leaves $54\,\%$ within the contour,
$84\,\%$ with $2\Delta\ln{\cal L}<6.0$, and $100\,\%$ with
$2\Delta\ln{\cal L}<9.2$. However, for power spectrum shape parameters
$\Omega_mh$ and $\Omega_b/\Omega_m$, $77\,\%$ have $2\Delta\ln{\cal
L}<2.3$, all but one have $2\Delta\ln{\cal L}<6.0$, and this mock has
$2\Delta\ln{\cal L}<9.2$.

Given the limited number of simulated catalogues, this is in
satisfactory agreement. We note that the mocks were drawn from the
Hubble Volume simulation (Evrard et~al. 2002), and are consequently
not completely independent. However, given that the numbers of mocks
within the expected confidence intervals are close to those expected
for independent Gaussian random variables, we feel justified in using
the standard $\chi^2$ intervals for our quoted parameters.

\section{tests of the method}  \label{sec:tests}

\begin{figure}
  \setlength{\epsfxsize}{1.0\columnwidth} 
    \centerline{\epsfbox{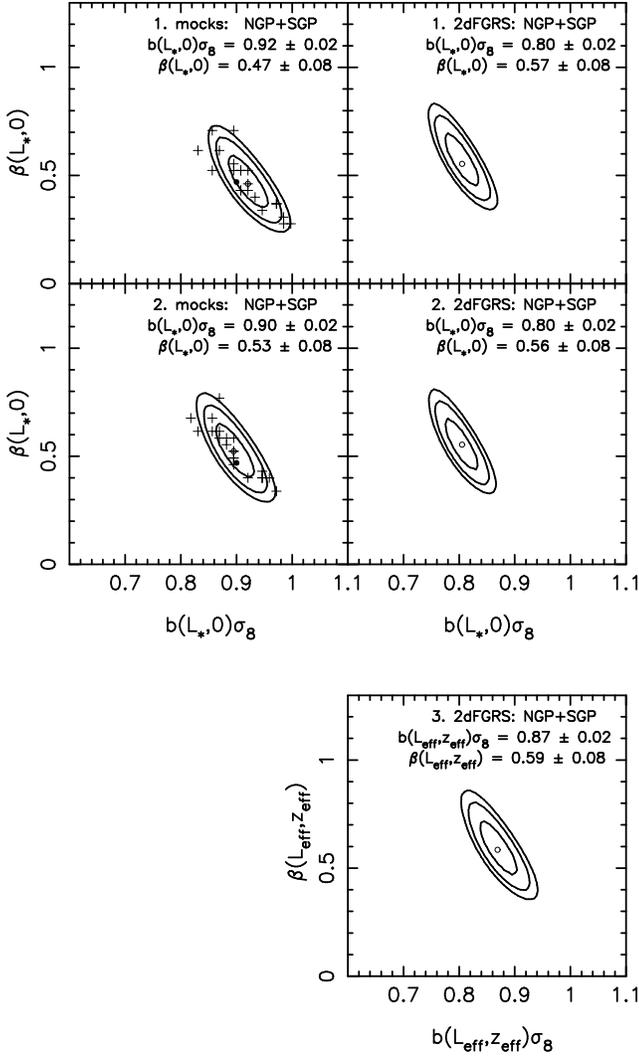}}

  \caption{Likelihood contour plots as in Fig.~\ref{fig:sig8_beta},
  but now designed to test the Spherical Harmonics method. The
  different rows correspond to models with: (1.)~power spectrum shape
  corresponding to linear rather than non-linear
  model. (2.)~1/completeness weighting for galaxies so that the
  weighted angular mask is uniform over the area of the
  survey. (3.)~No luminosity-bias correction. \label{fig:mod_test}}
\end{figure}

\begin{figure}
  \setlength{\epsfxsize}{1.0\columnwidth} 
    \centerline{\epsfbox{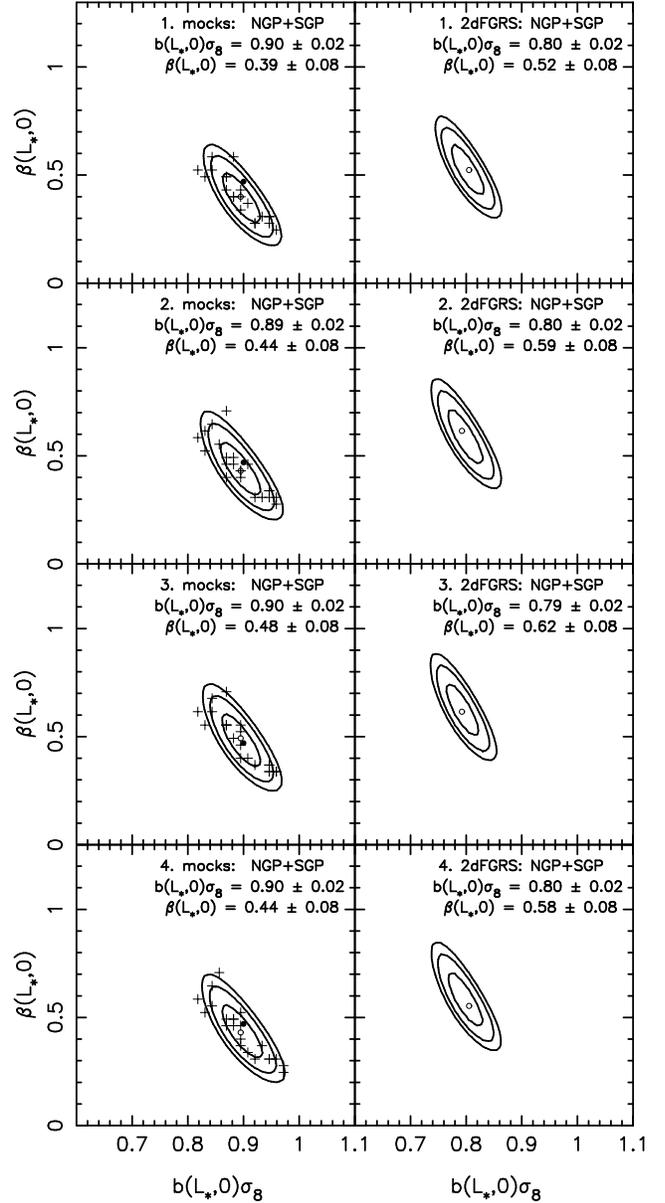}}

  \caption{Likelihood contour plots as in Fig.~\ref{fig:sig8_beta},
  but now designed to test the effect of the Fingers-of-God correction
  applied. The different rows correspond to models with: (1.)~no
  Fingers-of-God correction applied. (2.)~an exponential model for the
  probability distribution caused by the Fingers-of-God effect
  (Eq.~\ref{eq:Sexp}) with $\sigma_{\rm pair}=400\kms$. (3.)~a
  Gaussian model the probability distribution (Eq.~\ref{eq:Sgauss})
  with $\sigma_{\rm pair}=400\kms$. (4.)~a model with exponential
  distribution for the correlation function (Eq.~\ref{eq:Spair}) with
  $\sigma_{\rm pair}=400\kms$. \label{fig:ssvd_test}}
\end{figure}

In the Fourier based analysis of P01 and C04, the expected variation
in the measured power is only weakly dependent on the cosmological
parameters and a fixed covariance matrix could therefore be
assumed. In the analysis presented in this paper, the transformed
density field is modelled rather than the power, and the likelihood
variation due to cosmology is completely modelled using the covariance
matrix -- indeed, it is the variation of the covariance matrix that
alters the likelihood and allows us to estimate the cosmological
parameters. Consequently, we need to perform an inversion of this
matrix for each cosmological model to be tested (an $N^3$
operation). For a fixed power spectrum shape, the variation in the
inverse covariance matrix with $b(L_*,0)\sigma_8$ and $\beta(L_*,0)$
is small and we can use a iterative trick (described in
Section~\ref{sec:like_calc}) to estimate the covariance matrix using
an $N^2$ operation. The covariance matrix obviously varies more
significantly when we allow the cosmological parameters to vary more
freely and the shape of the power spectrum changes. A full matrix
inversion is then required for each model tested. This is
computationally intensive and consequently the specific tests
presented in this Section are based around recovering
$b(L_*,0)\sigma_8$ and $\beta(L_*,0)$ for a fixed $P(k)$ shape.

In Figs.~\ref{fig:mod_test} \&~\ref{fig:ssvd_test} we present
recovered likelihood surfaces calculated with various changes to our
method. These plots demonstrate tests of our basic assumptions and of
our implementation of the Spherical Harmonics method.

\subsection{mock catalogues}

In addition to the 2dFGRS results presented in
Figs.~\ref{fig:sig8_beta}, \ref{fig:omh_bf}
\&~\ref{fig:sig8_beta_marg}, we also plot contours revealing the
average likelihood surface determined from the 22 mock catalogues
described in Section~\ref{sec:mockcat}. The average surface is used so
the 2dFGRS and mock contours are directly comparable. For
$b(L_*,0)\sigma_8$ and $\beta(L_*,0)$ we also give the recovered
parameters and errors from the average likelihood surface. These
numbers can be compared with the expected values
$b(L_*,0)\sigma_8=0.9$ and $\beta(L_*,0)=0.47$. Crosses in these plots
show the maximum likelihood positions calculated from each of the mock
catalogues, while the open circle gives the combined maximum
likelihood position, and the solid circles shows the expected
values. We see that the recovered value of $\beta(L_*,0)$ is slightly
higher than expected. However, we will show in Section~\ref{sec:Btest}
that the recovered value of $\beta(L_*,0)$ changes in a consistent way
following changes in the peculiar velocities calculated for the
galaxies in each mock. Furthermore, we show that there is no evidence
for a systematic offset in the recovered $\beta(L_*,0)$, which we
would expect to vary with the peculiar velocities. The true value of
$b(L_*,0)\sigma_8$ is recovered to sufficient precision.

\subsection{Non-linear power assumption}

Although the width of the window function means that the modes are
dependent on the real-space power spectrum at $k>0.2\hompc$, this
dependence is weak compared with the dependence on the low-$k$, linear
regime (this is shown in Fig.~\ref{fig:win}). The likelihood
calculation used in Eq.~\ref{eq:like} relied on the transformed
density field having Gaussian statistics. While this is expected to be
true in the linear regime, on the scales of non-linear collapse this
assumption must break down. Although the modes must deviate from
Gaussianity, we consider the change in shape of the power as a first
approximation, and use the fitting formulae of Smith et~al.  (2003) to
determine the model power. In order to test the significance of this,
we consider the effect of replacing the non-linear power in the model
(Eq.~\ref{eq:cov_geom}) with the linear power. In fact, this has a
relatively small effect on the recovered power spectrum amplitude and
$\beta(L_*,0)$ as shown in Fig.~\ref{fig:mod_test}.

\subsection{Completeness weighting}

We have tried two angular weighting schemes for the galaxies. The
default weights do not have an angular component, and are simply the
radial weights of Feldman, Kaiser \& Peacock (1994) given by
Eq.~\ref{eq:wght}. For comparison we have also tried additionally
weighting each galaxy by 1/(angular completeness), so the weighted
galaxy density field at a given $r$ is independent of angular position
(i.e. it is uniform over the survey area). This weighting simplifies
the convolution of the model power to correct for the angular geometry
of the survey (Eq.~\ref{eq:D1} \&~\ref{eq:W}), and comparing results
from both schemes therefore tests this convolution. The downside of
such a weighting is the slight increase in shot noise. Results
calculated with this weighting scheme are presented in
Fig.~\ref{fig:mod_test}, and can be compared with the default in
Fig.~\ref{fig:sig8_beta}: no significant difference is observed
between the two schemes.

\subsection{Luminosity-dependent bias}  \label{sec:lumdep}

As described in Section~\ref{sec:bias_model}, we adopt a constant
galaxy clustering (CGC) model for the evolution of the fluctuation
amplitudes across the survey, and correct for the expected luminosity
dependence of this amplitude by weighting each galaxy by the
reciprocal of the expected bias ratio to $L_*$ galaxies (as suggested
by Percival, Verde \& Peacock 2004). The expected bias ratio assumed,
given by Eq.~\ref{eq:bobstar}, was calculated from a volume-limited
subsample of the 2dFGRS by Norberg et~al. (2001). In
Fig.~\ref{fig:mod_test}, we fit models that do not include either this
luminosity-dependent bias correction, or the evolution correction for
$\beta(L,z)$. This likelihood fit measures $\beta(L_{\rm eff},z_{\rm
eff})$, which is now a function of the effective luminosity $L_{\rm
eff}$ and effective redshift $z_{\rm eff}$ of the survey. For the
complete survey, examining the weighted density field gives that
$L_{\rm eff}=1.9L_*$ and $z_{\rm eff}=0.17$. However, we cannot be
sure that the Spherical Harmonics modes selected will not change these
numbers.

In fact, fitting to the data gives $\beta(L_{\rm eff},z_{\rm eff}) =
0.59\pm0.08$ and $b(L_{\rm eff},z_{\rm eff})\sigma_8(z_{\rm
eff})=0.87\pm0.02$. In order to compare these values with our results
that have been corrected for luminosity-dependent bias, we have to
consider a number of factors. For the CGC model,the change in the
measured power spectrum amplitude should only arise from the galaxy
luminosity probed. The effective luminosity of the sample is
$\sim1.9L_*$, which gives an expected bias of $1.13$ (using
Eq.~\ref{eq:bobstar}). The observed offset in amplitude is $1.08$,
perhaps indicating that, for the chosen modes, $L_{\rm eff}<1.9L_*$.
Within the CGC model, $\beta(L_{\rm eff},z_{\rm eff})$ is expected to
be related to $\beta(L_*,0)$ by
\be
  \beta(L_{\rm eff},z_{\rm eff}) =
    \frac{\Omega_m(z_{\rm eff})^{0.6}}{\Omega_m(0)^{0.6}}
    D(z_{\rm eff})
    \frac{b(L_*,0)}{b(L_{\rm eff},0)}
    \beta(L_*,0),
\ee
which gives $\beta(1.9L_*,0.17)\sim\beta(L_*,0)$ for a concordance
cosmological model as the different factors approximately cancel. In
fact, we measure no significant difference between $\beta(L_{\rm
eff},z_{\rm eff})$  and $\beta(L_*,0)$.

\subsection{Fingers-of-God correction} \label{sec:ssvd}

In this Section we test the assumed scattering probability that
corrects distance errors induced by the peculiar velocities of
galaxies inside groups. This probability was used to convolve the
model transformed density fields using the matrix presented in
Eq.~\ref{eq:S}. We compare models with exponential and Gaussian forms,
and a model that corresponds to an exponential convolution for the
correlation function (this corresponds to the model advocated by
Ballinger, Peacock \& Heavens 1996; Hawkins et~al. 2003)
\begin{eqnarray}
  p_e(r-y) &=& \frac{1}{\sqrt{2}\sigma_v}
    \exp\left[-\frac{\sqrt{2}\,|r-y|}{\sigma_v}\right], 
    \label{eq:Sexp}\\
  p_g(r-y) &=& \frac{1}{\sqrt{2\pi}\sigma_v}
    \exp\left[-\frac{(r-y)^2}{2\sigma_v^2}\right],
    \label{eq:Sgauss}\\
  p_b(r-y) &=& \frac{2\sqrt{2}}{\sigma_v}
    K_0\left[-\frac{\sqrt{2}}{\sigma_v}(r-y)\right].
    \label{eq:Spair}
\end{eqnarray}
$\sigma_v$ is the one-dimensional velocity dispersion, related to the
commonly used pairwise velocity dispersion by $\sigma_{\rm
pair}=\sqrt{2}\sigma_v$. $K_n$ is an $n$th-order modified Bessel
function derived as the inverse Fourier transform of the root of a
Lorentzian (Taylor et~al. 2001).

The Fingers-of-God effect stretches structure along the line-of-sight,
whereas large-scale bulk motions tend to foreshorten objects. Although
these effects predominantly occur on different scales, there is some
overlap, and if the Fingers-of-God effect is not included when
modelling the data, the best-fit value of $\beta(L_*,0)$ is decreased
slightly. In this case, the best-fit model interpolates between the
two effects, as demonstrated in Fig.~\ref{fig:ssvd_test}, where we
present the best-fit $\beta(L_*,0)$ with and without including the
Fingers-of-God correction.

In the results presented in Fig.~\ref{fig:sig8_beta}, we assumed an
exponential distribution for the distribution function of random
motions with $\sigma_{\rm pair}=350\kms$ for the 2dFGRS catalogue and
$\sigma_{\rm pair}=500\kms$ for the mock catalogues. We have tried a number of
different values of $0<\sigma_{\rm pair}<500\kms$ and find only very small
variation in the best-fit $\beta(L_*,0)$, as expected because we have
chosen modes that peak for $k<0.15\hompc$, where the finger-of-god
correction is small. To demonstrate this, in Fig.~\ref{fig:ssvd_test}
we present results calculated using Eqns.~\ref{eq:Sexp},
\ref{eq:Sgauss} \&~\ref{eq:Spair} with $\sigma_{\rm pair}=400\kms$ for both the
2dFGRS and mock catalogues. We also compare with the effect of not
including any correction for the small-scale velocity
dispersion. Little difference is seen in the recovered values of
$\beta(L_*,0)$, adding weight to the hypothesis that the
Fingers-of-God correction is not important for our determination of
$\beta(L_*,0)$.

In the $\xi(\sigma,\pi)$ analyses of the 2dFGRS presented in Peacock
et al. (2001), Madgwick et al. (2003) and Hawkins et al. (2003) a
strong degeneracy was revealed between the Fingers-of-God and linear
redshift-space distortions. Although such a degeneracy is also present
in the results from our Spherical Harmonics analysis, it is weak
compared with the $\xi(\sigma,\pi)$ results. The difference is due to
the scales analysed -- the correlation function studies estimated the
clustering strength on smaller scales where the Fingers-of-God
convolution is more important. Because the Fingers-of-God effect has
less effect in our analysis, we are less able to constrain its
amplitude, and therefore assume a fixed value motivated by the
$\xi(\sigma,\pi)$ analyses, rather than fitting to the data.

\subsection{2dFGRS catalogue calibration} \label{sec:calib}

\begin{figure*}
  \setlength{\epsfxsize}{0.7\textwidth} 
    \centerline{\epsfbox{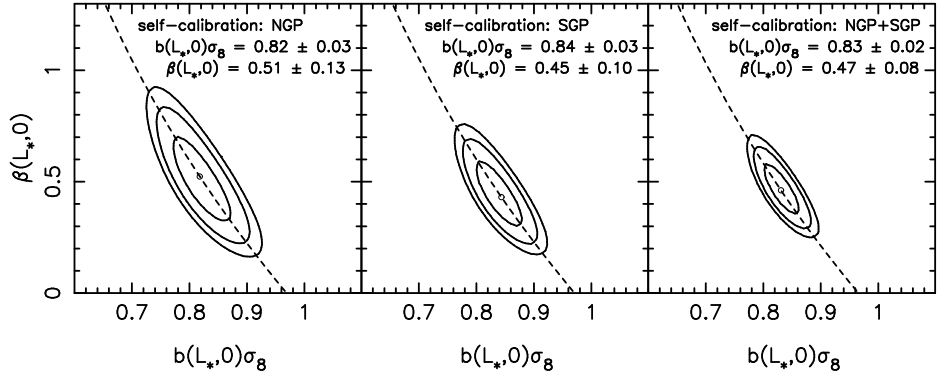}}

  \caption{Likelihood contours for the recovered $b(L_*,0)\sigma_8$
  and $\beta(L_*,0)$ assuming a fixed $\Lambda$CDM power spectrum
  shape as in Fig.~\ref{fig:sig8_beta}, but now calculated having
  revised the calibration of the 2dFGRS catalogue. Details of the
  revised calibration are presented in Section~\ref{sec:calib}.
  \label{fig:sig8_beta_self}}
\end{figure*}

Outwith this Section we consider the 2dFGRS release catalogue and
corresponding calibration. In order to test the dependence of our
results on the catalogue calibration, in this Section we report on the
analysis of a different version of the catalogue with revised
calibration. In the release catalogue (Colless et~al. 2003; {\tt
http://www.mso.anu.edu.au/2dFGRS/}) the 2dFGRS photographic magnitudes
were calibrated using external CCD data from the SDSS Early Data
Release and ESO Imaging Survey (EIS) (Colless et~al. 2003; Cross et
al. 2003). Overlaps between the photographic plates allow this
calibration to be propagated to the whole survey. In this section we
instead calibrate each plate without the use of external data. The
magnitudes in the final released catalogue, $b_{\scriptscriptstyle\rm
J}^{\rm final}$ and magnitudes, $b_{\scriptscriptstyle\rm J}^{\rm
self}$, resulting from this self-calibration are assumed to be related
by a linear relation
\be
  b_{\scriptscriptstyle\rm J}^{\rm self} =  a_{\rm self}  
    b_{\scriptscriptstyle\rm J}^{\rm final} + b_{\rm self},
\ee
where the calibration coefficients $a_{\rm self}$ and $b_{\rm self}$
are allowed to vary from plate to plate. To set the values of these
calibration coefficients two constraints are applied. First on each
plate we assume that the galaxy luminosity function can be represented
by a Schechter function with faint-end slope $\alpha=1.2$ and make a
maximum likelihood estimate of $M_*$.  The value of $M_*$ is sensitive
to the difference in $b_{\scriptscriptstyle\rm J}^{\rm self}$ and
$b_{\scriptscriptstyle\rm J}^{\rm final}$ at around
$b_{\scriptscriptstyle\rm J}=17.5$ and the number of galaxies on each
plate is such that the typical random error on $M_*$ is $0.03$
magnitudes.  Second we compare the number of galaxies, $N(z>0.25)$,
with redshifts greater than $z=0.25$ with the number we expect,
$N_{\rm model}(z>0.25)$, based on our standard model of the survey
selection function.  The value of $N_{\rm model}(z>0.25)$ depends
sensitively on the survey magnitude limit and so constrains the
difference in $b_{\scriptscriptstyle\rm J}^{\rm self}$ and
$b_{\scriptscriptstyle\rm J}^{\rm final}$ at $b_{\scriptscriptstyle\rm
J} \approx 19.5$. By demanding that on each plate both $N(z>0.25)
=N_{\rm model}(z>0.25)$ and $M_*-5 \log h=19.73$ we determine $a_{\rm
self}$ and $b_{\rm self}$.  This method of calibrating the catalogue
is extreme in that it ignores the CCD calibrating data (apart from
setting the overall zero point of $M_*-5 \log h=19.73$). A more
conservative approach is to combine the external calibration with the
internal one and determine $a_{\rm self}$ and $b_{\rm self}$ by a
$\chi^2$ procedure that takes account of the statistical error on
$M_*$, the expected variance on $N(z>0.25)$ given by mock catalogues
and the errors on the calibrating data. Unless the errors on the CCD
calibration are artificially inflated this results in a calibration
very close to that of final release.  Thus we believe that the
difference between the results achieved with self-calibrated catalogue
and the standard final catalogue represent an upper limit on the
effects attributable to uncertainty in the photometric calibration.

For the Spherical Harmonics analysis method, we need to cut the 2dFGRS
catalogue so that the radial distribution of galaxies is independent
of angular position (this catalogue reduction was described in
Section~\ref{sec:2dFcat}). Changing the magnitude limit at which to
cut the catalogue changes the angular mask for the reduced sample as
angular regions that do not go as faint as the chosen limit are
removed. Rather than optimize the magnitude limit at which to cut the
self-calibrated catalogue, we instead resample the revised catalogue
using the old mask. A magnitude limit was then chosen to fully sample
this angular region and give a radial distribution that is independent
of angular position. This procedure avoided the computationally
expensive recalculation of angular matrices (see
Appendix~\ref{app:A}). However, the radial galaxy distribution and
total number of galaxies were different from those in our primary
analysis, and a revised radial component of the covariance matrix was
required.

Revised parameter constraints on $\beta(L_*,0)$ and $b(L_*,0)\sigma_8$
are presented in Fig.~\ref{fig:sig8_beta_self}, which can be directly
compared with the upper panels in Fig.~\ref{fig:sig8_beta}. An
incorrect calibration would lead to a resampling of the complete
2dFGRS catalogue (described in Section~\ref{sec:2dFcat}) that would
not produce a catalogue with radial galaxy distribution independent of
angular position. This would lead to an increase in the amplitude of
the observed angular fluctuations. Given that we split the
fluctuations into an overall power spectrum and an additional
component in the radial direction caused by linear redshift space
distortions, an artificial increase in angular clustering would
manifest itself as an increase in $b(L_*,0)\sigma_8$, coupled with a
decrease in $\beta(L_*,0)b(L_*,0)\sigma_8$, which controls the
absolute amplitude of the linear redshift-space distortions. In fact
this is exactly what is observed when comparing results from the
release and self-calibration catalogues (Figs.~\ref{fig:sig8_beta}
\&~\ref{fig:sig8_beta_self}), suggesting that the self-calibration
procedure introduces artificial angular distortions into the reduced
catalogue. The dashed lines in Fig.~\ref{fig:sig8_beta_self} show the
locus of models with redshift-space power spectrum amplitude
(calculated from Eq.~\ref{eq:zspacepower}) at the maximum likelihood
(ML) value. Comparing the relative positions of the ML points in
Figs.~\ref{fig:sig8_beta} \&~\ref{fig:sig8_beta_self} shows that
changing the catalogue calibration moves the maximum likelihood
position along this locus, without significantly changing the
redshift-space power amplitude. Catalogue calibration and selection
represents the most significant potential source of systematic error
in our analysis.

\subsection{Testing $\beta(L_*,0)$ using mock catalogues}  \label{sec:Btest}

\begin{figure}
  \setlength{\epsfxsize}{0.65\columnwidth}
    \centerline{\epsfbox{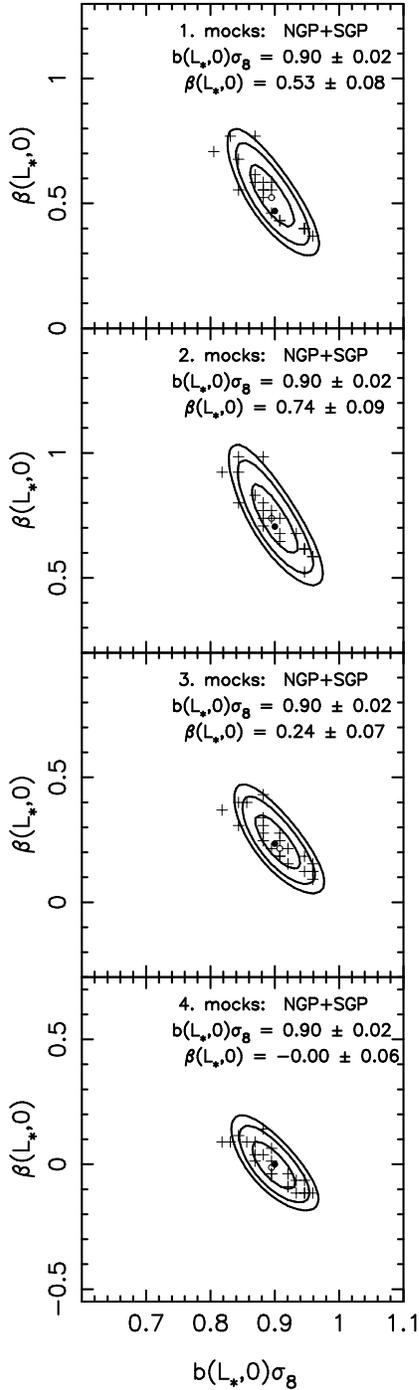}}

  \caption{Likelihood contour plots as in Fig.~\ref{fig:sig8_beta},
  but now designed to test our recovery of $\beta(L_*,0)$ using the
  mock catalogues. The different panels correspond to: 1.~the standard
  catalogues with peculiar velocities calculated directly from the
  Hubble Volume simulation. Here the true value of $\beta(L_*,0)$ is
  $0.47$. 2.~the contribution to the galaxy redshifts from the
  peculiar velocities has been increased by 50\,\%. Here, we assume
  $\sigma_{\rm pair}=750\kms$. Without this correction, $\beta(L_*,0)$
  would increase by less than 50\,\%. We expect $\beta(L_*,0)=0.71$,
  shown by the solid circle.  3.~as 2, but now decreasing the redshift
  contribution by 50\,\%. $\sigma_{\rm pair}=250\kms$ is assumed, and
  we expect $\beta(L_*,0)=0.24$. 4.~recovered parameters from real
  space catalogues, calculated assuming that $\sigma_{\rm
  pair}=0$. Obviously, we expect to recover
  $\beta(L_*,0)=0.0$. \label{fig:Btest}}
\end{figure}

For each galaxy in the mock catalogues, we know the relative
contributions to the redshift from the Hubble flow and peculiar
velocity. Consequently, we can easily increase or decrease the
amplitude of the redshift space distortions to mimic catalogues with
different cosmological parameters. In Fig.~\ref{fig:Btest} we plot the
recovered power spectrum amplitudes and $\beta(L_*,0)$ from catalogues
created by increasing or decreasing the peculiar velocity by 50\,\%
from the true value. Obviously, this changes both the linear redshift
space distortions and the Fingers-of-God, and consequently we fit to
these data assuming a revised $\sigma_{\rm pair}$. If we neglected to
do this, the average recovered $\beta(L_*,0)$ would vary from the true
value by less than 50\,\%, as assuming the wrong value of $\sigma_{\rm
pair}$ has the effect of damping the change in the recovered
$\beta(L_*,0)$. However, because the linear redshift space distortions
are dominant on the scales being probed in our analysis, we would
still see a change in the correct direction. For these catalogues, we
find that altering the peculiar velocities results in a consistent
change in the recovered $\beta(L_*,0)$, showing that our likelihood
test is working well.

In Fig.~\ref{fig:Btest} we also show the recovered parameter
constraints from mocks catalogues with no redshift space
distortions. Here, we fitted to these data assuming that there was no
Fingers-of-God effect, and see that we recover for $\beta(L_*,0)$
consistent with $0$ for each catalogue.

\section{summary and discussion}  \label{sec:discussion}

\begin{figure}
  \setlength{\epsfxsize}{0.65\columnwidth} 
    \centerline{\epsfbox{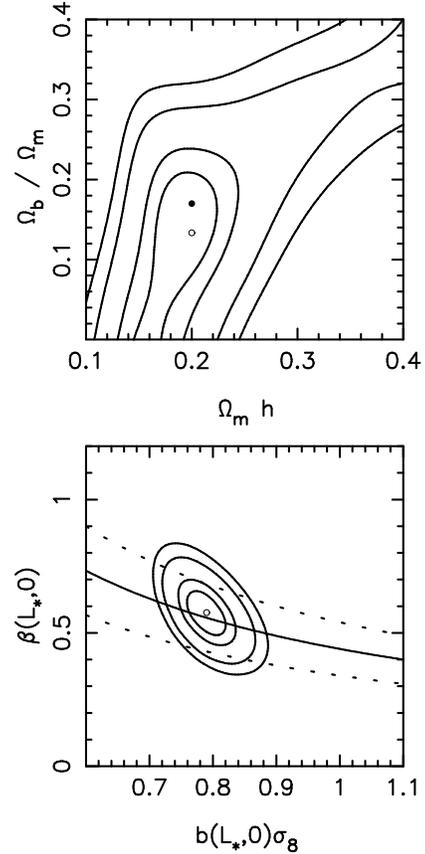}}

  \caption{Likelihood contour plots for the combined NGP + SGP 2dFGRS
  catalogue as in Fig.~\ref{fig:param_degen} compared with best fit
  parameters from WMAP (Bennett et~al. 2003; Spergel et~al. 2003). The
  constraint on the characteristic amplitude of velocity fluctuations
  from the 1-year WMAP data is $\sigma_8\Omega_m^{0.6}=0.44\pm0.10$,
  which is shown in the left panel by the thick solid line, with
  1$\sigma$ errors given by the dotted lines. In the right panel, the
  solid circle shows the best-fit parameter values of $\Omega_mh=0.20$
  \& $\Omega_b/\Omega_m=0.17$. As can be seen, the constraints
  resulting from the 2dFGRS power spectrum shape and the linear
  distortions are consistent with the WMAP
  data. \label{fig:wmap_comparison}}
\end{figure}

The Spherical Harmonics analysis method of HT and Tadros et~al. (1999)
has been extended and updated to allow for surveys that cover a
relatively small fraction of the sky. Additionally, a consistent
approach has been adopted to model luminosity-dependent bias and the
evolution of the matter power spectrum. We assume a constant galaxy
clustering model for the redshift region $0<z<0.25$ covered by the
2dFGRS survey, in which, although the matter density field does
evolve, the galaxy power spectrum is assumed to remain fixed. Galaxy
bias is also assumed to be a function of luminosity, and we correct
for the effect that this has on the recovered power spectrum.

The revised method has been applied to the complete 2dFGRS catalogue,
resulting in tight constraints on the amplitude of the linear redshift
space distortions. Because the method still requires a survey with
selection function separable in radial and angular directions, we have
to use a reduced version of the final 2dFGRS catalogue. Additionally,
we are forced to compromise on the quantity of data (number of modes)
analysed, although we have tried to perform a logical and optimized
reduction of the mode number. These considerations mean that we do not
obtain the accuracy of the cosmological constraints from the shape of
the galaxy power spectrum obtained in our companion Fourier analysis
(C04). This reduction in accuracy primarily results from the decrease
in the catalogue size. In particular, the analysis is cosmic variance
limited and most of the discarded galaxies were luminous and therefore
at high redshift where they trace a large volume of the
Universe. However, from the Spherical Harmonics method we do obtain
power spectrum shape constraints $\Omega_b/\Omega_m<0.21$ as shown in
Fig.~\ref{fig:omh_bf} and, for fixed $\Omega_b/\Omega_m=0.17$, we find
$\Omega_mh=0.20^{+0.03}_{-0.03}$, consistent with previous power
spectrum analyses from the 2dFGRS and the SDSS.

We have also modelled the overdensity distribution in 22 LCDM mock
catalogues, designed to mimic the 2dFGRS. By presenting recovered
parameters from these catalogues, we have shown that any systematic
biases induced by the analysis method are at a level well below the
cosmic variance caused by the size of the survey volume. In
particular, it should be emphasized that these mocks include a
realistic degree of scale-dependent bias, to reflect the known
difference in small-scale clustering between galaxies and the
nonlinear CDM distribution (e.g. Jenkins et al. 1998). We have
additionally used these catalogues to test the errors placed on
recovered parameters and find that assuming a $\chi^2$ distribution
for $\ln{\cal L}$ provides approximately the correct errors. The tests
presented, considering the NGP and SGP separately, using the mocks,
and varying parts the analysis method were designed to test our
Spherical Harmonics formalism and the assumptions that go into
this. In particular, we do not test the 2dFGRS sample for internal
consistency, for instance splitting by redshift or magnitude limit,
although we do find consistent parameter estimates from the northern
and southern parts of the survey. A more comprehensive set of tests is
presented in C04, using Fourier methods to decompose the density
field.

By considering a revised 2dFGRS catalogue calibration, we have
examined the effect of small systematic magnitude errors on our
analysis. Such errors artificially increase the strength of the
angular clustering, leading to an increase in the best-fit
$b(L_*,0)\sigma_8$ and a corresponding decrease in $\beta(L_*,0)$. We
have shown that the revised catalogue tested produces such a change in
the recovered parameters, therefore providing evidence in favour of
the release calibration. The calibration method and its effect will be
further discussed in C04. Here, we simply note that the systematic
error in $\beta(L_*,0)$ and $b(L_*,0)$ from catalogue calibration is
of the same order as the random error.

The strength of the Spherical Harmonics method as applied to the
2dFGRS lies in measuring the linear redshift-space distortions, and
fitting the real-space power spectrum amplitude. Consequently we are
able to measure $\beta(L_*,0)=0.58\pm0.08$, and
$b(L_*,0)\sigma_8=0.79\pm0.03$ for $L_*$ galaxies at $z=0$,
marginalising over the power spectrum shape. This result is dependent
on the constant galaxy clustering model and on the bias-luminosity
relationship derived by Norberg et~al. (2001), and covers
$0.02<k<0.15\hompc$. Our measurement of $\beta(L_*,0)$ is derived on
larger scales than the $\xi(\sigma,\pi)$ analyses of the 2dFGRS
presented in Peacock et al. (2001) and Hawkins et al. (2003), and
scale-dependent bias could therefore explain why our result is
slightly higher than the numbers obtained in these analyses.

Tegmark, Hamilton \& Xu (2002) performed a similar spherical harmonics
analysis of the 100k release of the 2dFGRS. As in the analysis
presented here, they also required a catalogue that was separable in
radial and angular directions, and cut the 100k release catalogue to
66050 galaxies. From these galaxies, they measured $\beta(L_{\rm
eff},z_{\rm eff})=0.49\pm0.16$, consistent with our result (see
Section~\ref{sec:lumdep} for a discussion of the conversion between
$\beta(L_{\rm eff},z_{\rm eff})$ and $\beta(L_*,0)$). Our result not
only allows for luminosity-dependent bias and evolution, it also uses
over twice as many galaxies as the Tegmark, Hamilton \& Xu (2002)
analysis.

On large-scales, galaxies are expected to directly trace the bulk
motion of the density field, so the absolute amplitude of the observed
redshift-space distortions caused by this motion is expected to be
independent of galaxy properties. This assumption has been tested
empirically by considering the mean relative velocity of galaxy pairs
in different samples (Juszkiewicz et~al. 2000; Feldman
et~al. 2003). Rather than fitting $\beta(L_*,0)$, the relative
importance of the linear redshift-space distortions compared with the
real-space galaxy power spectrum, we can instead fit the absolute
amplitude of these fluctuations. This results in the cosmological
constraint $\Omega_m^{0.6}\sigma_8=0.46\pm0.06$.

The relatively high power of $\sigma_8$ compared to $\Omega_m$ in this
constraint means that an additional constraint on $\Omega_m$ provides
a tight constraint on $\sigma_8$. For example, fixing $\Omega_m=0.3$
gives $\sigma_8=0.95\pm0.12$ ($\sim15$\% error), while fixing
$\sigma_8=0.95$ gives $\Omega_m=0.3\pm0.08$ ($\sim26$\% error). We
note that our constraint is approximately 1$\sigma$ higher than a
recent combination of weak-lensing measurements that gave
$\sigma_8\simeq0.83\pm0.04$ for $\Omega_m=0.3$ (Refregier
2003). Additionally, combining the weak-lensing constraint with our
measurement of $b(L_*,0)\sigma_8$ suggests that $b(L_*,0)\sim0.9$ in
agreement with the analyses of Lahav et~al. (2002) \& Verde
et~al. (2002). Similarly, combining our measurement of $\beta(L_*,0)$
with recent constraints on $\Omega_m$ (such as those derived by
Spergel et~al. 2003) suggests that $b(L_*,0)\sim0.9$, and we see that
we have a consistent picture of both the amplitude of the real-space
power spectrum and linear redshift-space distortions within the
concordance $\Lambda$CDM model.

A comparison of our results with parameter constraints from WMAP is
presented in Fig.~\ref{fig:wmap_comparison}. In this paper, we do not
attempt to perform a full likelihood search for the best-fit
cosmological model using the combined 2dFGRS and WMAP data
sets. Instead, we simply consider the consistency between the WMAP
data and our measurements of the 2dFGRS. In
Fig.~\ref{fig:wmap_comparison} we plot the WMAP constraint on
$\Omega_m^{0.6}\sigma_8$ as derived in Spergel et~al. (2003), compared
with our constraints on $\beta(L_*,0)$ and $b(L_*,0)\sigma_8$. WMAP
obviously tells us nothing about $b(L_*,0)$, so there is a perfect
degeneracy between these parameters. However, the constraints are seen
to be consistent. In fact our constraint is a significant improvement
on the WMAP constraint, primarily because of the uncertainty in the
optical depth to the last scattering surface, parameterized by $\tau$.

Because $h=0.72$ is fixed in the simple cosmological model assumed to
parameterise the power spectrum shape, the horizon angle degeneracy
for flat cosmological models (Percival et~al 2002; Page et~al. 2003)
is automatically lifted. The position of the first peak in the CMB
power spectrum therefore provides a tight constraint on $\Omega_m$.In
fact, given this simple model, the constraints on $\Omega_m$ and
$\Omega_b/\Omega_m$ from WMAP are so tight that we chose to plot a
point to show them in Fig.~\ref{fig:wmap_comparison}, rather than a
confidence region. However, had we considered a larger set of models
in which $h$ was allowed to vary, then an extra constraint is required
to break the horizon angle degeneracy even for the WMAP data (Page
et~al. 2003). In this paper we provide a new cosmological constraint
by measuring the strength of the linear distortions caused by the bulk
flow of the density field mapped by the final 2dFGRS catalogue.

\section*{ACKNOWLEDGMENTS}

The 2dF Galaxy Redshift Survey was made possible through the dedicated
efforts of the staff of the Anglo-Australian Observatory, both in
creating the 2dF instrument and in supporting it on the telescope. WJP
is supported by PPARC through a Postdoctoral Fellowship. JAP and OL
are grateful for the support of PPARC Senior Research Fellowships.

\appendix

\section{Method}  \label{app:A}

The Spherical Harmonics method applied to the 2dFGRS in this paper has
a number of significant differences from the formalisms developed for
the IRAS surveys (Fisher et~al. 1994;1995; HT; Tadros
et~al. 1999). The revisions are primarily due to the complicated
geometry of the 2dFGRS survey (whereas the IRAS surveys nearly covered
the whole sky), although we additionally apply a correction for
varying galaxy bias, dependent on both galaxy luminosity and
redshift. For these reasons we provide a simple, complete description
of the formalism in this appendix. Note that throughout we use a
single Greek subscript (e.g. $\nu$) to represent a triplet
(e.g. $\ell_\nu$,$n_\nu$,$m_\nu$), so the spherical harmonic
$Y_\nu(\theta,\phi)\equiv Y_{\ell_\nu m_\nu}(\theta,\phi)$ and the
spherical Bessel functions, $j_\nu(s)\equiv j_{\ell_\nu}(k_{\ell_\nu
n_\nu}s)$. $-\nu$ is defined to represent the triplet
($\ell_\nu$,$-m_\nu$,$n_\nu$). We also adopt the following convention
for coordinate positions: $\r$ is the true (or real space) position of
a galaxy, $\s$ is the observed redshift space position given the
linear in-fall velocity of the galaxy.  $\sp$ and $\rp$ correspond to
$\s$ and $\r$ including the systematic offset in the measured distance
caused by the small-scale velocity dispersion of galaxies within
larger virialised objects.

\subsection{Galaxy bias model}  \label{sec:bias_model}

As in Lahav et~al. (2002), we adopt a constant galaxy clustering (CGC)
model for the evolution of galaxy bias over the range of redshift
covered by the 2dFGRS sample used in this analysis
($0<z<0.25$) i.e. we assume that the normalization of the galaxy
density field is independent of redshift, for any galaxy luminosity
$L$. We also assume that the relative expected bias $\hat{r}_b(L)$
of galaxies of luminosity $L$ relative to that of $L_*$
galaxies is a function of luminosity
\be
  \hat{r}_b(L) = \llangle \frac{b(L,z)}{b(L_*,z)}\rrangle 
    = 0.85 + 0.15\frac{L}{L_*},
  \label{eq:bobstar} 
\ee 
and that this ratio is independent of redshift. This dependence is
implied by the relative clustering of 2dFGRS galaxies (Norberg
et~al. 2001).

In the analysis presented in this paper, the galaxy bias is modelled
using a very simple linear form with the mean redshift-space
density of galaxies of luminosity $L$ given by
\begin{eqnarray}
  \rho(\rp) & = & \bar{\rho}(\rp)\left[1 
    + b(L,0)\delta({\rm mass},\rp)\right]\\
  & = & \bar{\rho}(\rp)\left[1 
    + \hat{r}_b(L)\delta(L_*,\rp)\right],
  \label{eq:biaseddelta}
\end{eqnarray}	
where $\delta({\rm mass},\rp)$ is the present day mass density field,
and $\delta(L_*,\rp)$ is the density field of galaxies of luminosity
$L_*$, which is assumed to be independent of epoch.

The galaxy bias model described above was used to correct the observed
galaxy overdensity field, enabling measurement of the shape and
amplitude of the power spectrum of $L_*$ galaxies. Following the CGC
model, we do not have to correct the recovered clustering signal for
evolution, provided that we wish to measure the galaxy rather than the
mass power spectrum. However, because galaxy luminosity varies
systematically with redshift, we do need to correct for
luminosity-dependent bias, and we do this in a way analogous to the
Fourier method presented by Percival, Verde \& Peacock (2004), by
weighting the contribution of each galaxy to the measured density
field by the reciprocal of the expected bias ratio $\hat{r}_b(L)$
given by Eq.~\ref{eq:bobstar}.

In the following description of the formalism, we only consider
galaxies of luminosity $L$. Without loss of generality, this result
can be expanded to cover a sample of galaxies with different
luminosities by simply summing (or integrating) over the range of
luminosities (as in Percival, Verde \& Peacock 2004).

\subsection{The Spherical Harmonic formalism}

Further description of the Spherical Harmonics formalism may be found
in Fisher et al. (1994; 1995), HT and Tadros et~al. (1999). Expanding
the density field of the redshift-space distribution of galaxies of
luminosity $L$ in spherical harmonics and spherical Bessel functions
gives
\be
  \rho_\nu(L,s') = c_\nu \int\!d^3\!s' \,
    \frac{\rho(L,\sp)}{\hat{r}_b(L)} 
    w(\sp) j_\nu(s') Y^*_\nu(\theta,\phi),
    \label{eq:sht}
\ee
where $w(\sp)$ is a weighting function for which we adopt the standard
Feldman, Kaiser \& Peacock (1994) weight
\be
  w(\sp) = \frac{1}{1+\bar{\rho}(\sp)\langle P(k)\rangle}.
  \label{eq:wght}
\ee
Here, $\bar{\rho}(\sp)$ is the mean galaxy redshift-space density for
all galaxies, $\langle P(k)\rangle$ is an estimate of the power
spectrum, and $\sp$ is the 3D redshift-space position variable. Note
that, to simplify the procedure, we do not use luminosity-dependent
weights as advocated by Percival, Verde \& Peacock (2004). $c_\nu$ are
normalization constants, and $\rho(\sp)$ is the galaxy redshift-space
density. For a galaxy survey, $\rho(\sp)$ is a sum of delta functions
and the above integral decomposes to a sum over the galaxies.

The inverse transformation is given by
\be
  \frac{\rho(L,\sp)}{\hat{r}_b(L)}w(\sp) = \sum_\nu c_\nu \rho_\nu(L,s') 
    j_\nu(s') Y_\nu(\theta,\phi).
\ee
Adopting the set of harmonics with 
\be
 \left.\frac{d}{dr}j_\nu(r)\right|_{r_{\rm max}} = 0,
\ee
(i.e. with no boundary distortions at $r_{\rm max}=706.2\mpcoh$), the
normalization of the transform requires $c_\nu$ to satisfy
\be
 c^{-2}_\nu = \int\!dr\,j^2_\nu(r) r^2.
\ee

\subsection{Small-scale velocity dispersion correction}

We have applied the correction (described by HT) for the effect of the
small-scale non-linear peculiar velocity field caused by the random
motion of galaxies in groups. Because we are only interested in the
large-scale linear power spectrum in this paper, the exact details of
this correction are not significant (this is discussed further in
Section~\ref{sec:ssvd}). The effect of the velocity field is to smooth
the observed overdensity field along the line-of-sight in a way that
is equivalent to convolving with a matrix $S_{\nu\mu}$:
\be
  (\delta_{r'})_\nu = \sum_\mu S_{\nu\mu} (\delta_{r})_\mu,
  \label{eq:Sconv}
\ee
where
\bm 
  S_{\nu\mu} = c_\nu c_\mu \Delta^K_{\ell_\nu,\ell_\mu} 
    \Delta^{K}_{m_\nu,m_\mu} \;\times
\em
\be
  \hspace{1cm} \int\!\!\!\!\int p(r-y)
    j_\mu(r)j_\nu(y) \,r dr \,y dy.
    \label{eq:S} 
\ee
Here $\Delta^K$ is the Kronecker delta function, and $p(r-y)$ is the
one-dimensional scattering probability for the velocity
dispersion. Models for $p(r-y)$ are given in Eqns.~\ref{eq:Sexp},
\ref{eq:Sgauss}, \&~\ref{eq:Spair}, and the choice of model is
discussed further in Section~\ref{sec:ssvd}. Note that this formalism
assumes that the induced dispersion is not a strong function of group
mass. 

\subsection{Modelling the transformed density field}  
\label{sec:method_main}

The correction for luminosity-dependent bias given by
Eq.~\ref{eq:bobstar} is a function of galaxy properties, not the
measured galaxy position. The galaxy density multiplied by this bias
correction is therefore conserved with respect to a change in
coordinates with number conservation implying
\be
  d^3\sp\frac{\rho(L,\sp)}{\hat{r}_b(L)} = 
  d^3\rp\frac{\rho(L,\rp)}{\hat{r}_b(L)}.
\ee

The dependence of the redshift distortion term lies in the weighting
and spherical Bessel functions and, following HT, we expand to first
order in $\Delta r'\equiv s'-r'$,
\be
  w(\sp)j_\nu(s') \simeq w(\rp)j_\nu(r') + \Delta
    r'\frac{d}{dr}\left[w(\rp)j_\nu(r')\right].
    \label{eq:pv1}    
\ee

Using the Poisson equation to relate the gravitational potential with
the density field, 
\bm
  \Delta r'_{\rm lin} = \Omega_m(z[r'])^{0.6} \;\times
\em
\be
  \hspace{1cm}
    \sum_\nu  \frac{1}{k_\nu^2} c_\nu 
    \delta_\nu({\rm mass},\rp) \frac{dj_\nu(r')}{dr} Y_\nu(\theta,\phi),
\ee
where $\delta_\nu({\rm mass},r')$ is the transform of the mass
over-density field. Because the linear redshift-space distortions are
a function of the mass over-density field, they are independent of
galaxy luminosity. However, this means that they are expected to grow
through the linear growth factor $D(z)$, normalized to $D(0)=1$,
within the CGC model. This and the redshift dependence of
$\Omega_m(z)/\Omega_m(0)$ are calculated assuming a concordance
model. We can now rewrite these distortions in
terms of the transformed density field of galaxies of luminosity $L_*$
\bm
  \Delta r'_{\rm lin} = \frac{\Omega_m(z[r'])^{0.6}}{b(L_*,0)}
    D(z[r']) \;\times
\em
\be
  \hspace{1cm}
    \sum_\nu  \frac{1}{k_\nu^2} c_\nu 
    \delta_\nu(L_*,\rp) \frac{dj_\nu(r')}{dr} Y_\nu(\theta,\phi).
\ee
Defining 
\be
  \beta(L_*,0)\equiv\frac{\Omega_m(0)^{0.6}}{b(L_*,0)},
\ee 
reduces this expression to
\bm
  \Delta r'_{\rm lin} = \beta(L_*,0)
    \frac{\Omega_m(z[r'])^{0.6}}{\Omega_m(0)^{0.6}} D(z[r']) 
\em
\be
  \hspace{1cm}
    \sum_\nu  \frac{1}{k_\nu^2} c_\nu \delta_\nu(L_*,\rp)
    \frac{dj_\nu(r')}{dr} Y_\nu(\theta,\phi).
\ee
Including a correction for the local group velocity
$\vlg$, assumed to be $622\kms$ towards (B1950) ${\rm
RA}=162^\circ$, ${\rm Dec}=-27^\circ$ (Lineweaver et~al. 1996;
Courteau \& van den Bergh 1999), gives
\be
  \Delta r' = \Delta r'_{\rm lin} - \vlg\cdot\hrp.
  \label{eq:pv2}    
\ee
The local group velocity correction has a very minor effect on the
results presented in this paper, but was included for completeness.

For galaxies of luminosity $L$, transforming the density field gives 
\bm
  \rho(L,\rp) = \bar{\rho}(L,\rp) \;\times
\em
\be
  \hspace{1cm}
    \left[1 + \sum_\nu c_\nu 
    \delta_\nu(L,\rp) j_\nu(r') Y_\nu(\theta,\phi)\right],
    \label{eq:deltaL}
\ee
where $\bar{\rho}(L,\rp)$ is the observed mean density of galaxies of
luminosity $L$ in the survey. In fact, the mean number of galaxies as
a function of the redshift-space distance $\bar{\rho}(L,\sp)$ is more
easily determined than $\bar{\rho}(L,\rp)$. It would be possible to
reformulate the Spherical Harmonics formalism to use
$\bar{\rho}(L,\sp)$ by separating the convolution of the window from
the linear redshift-space distortion correction. Given the relatively
small effect that the coordinate translation $\rp\to\sp$ has on
$\bar{\rho}(L,\rp)$, we have instead chosen to use the original HT
formalism with $\bar{\rho}(L,\rp)\simeq\bar{\rho}(L,\sp)$ as measured
from the survey. Converting from $\delta_\nu(L,\rp)$ to consider the
fluctuations traced by $L_*$ galaxies gives
\bm
  \rho(L,\rp) = \bar{\rho}(L,\rp) \;\times
\em
\be
  \hspace{1cm}
    \left[1 + \sum_\nu c_\nu 
    \hat{r}_b(L)\delta_\nu(L_*,\rp) j_\nu(r') Y_\nu(\theta,\phi)\right],
    \label{eq:delta}
\ee
and we see that when we combine Eqns.~\ref{eq:sht}, \&~\ref{eq:delta}
to determine $\rho(L,\rp)$ as a function of
$\langle\delta_\nu(L_*,\rp)\rangle$, the factors of $\hat{r}_b(L)$ in
both of these Equations will cancel.

Combining Eqns.~\ref{eq:sht}, \ref{eq:Sconv},
\ref{eq:pv1}, \ref{eq:pv2}~\& \ref{eq:delta} gives 
\begin{eqnarray}
  D_\nu & \equiv & \rho_\nu(L,\rp) - \bar{\rho}_\nu(L,\rp) -
          \rho_\nu({\rm LG},\rp) \\
    & = & \sum_\mu\left(\Phi_{\nu\mu} + \beta(L_*,0)
          V_{\nu\mu}\right)\delta_\mu(L_*,\rp) \\
    & = & \sum_\mu\sum_\eta\left(\Phi_{\nu\mu} + \beta(L_*,0)
          V_{\nu\mu}\right)S_{\mu\eta}\delta_\eta(L_*,\r)
	    \label{eq:D1}
\end{eqnarray}	
where the mean-field harmonics are defined as
\be
  \bar{\rho}_\nu(L,\rp) = c_\nu \int\!d^3\!r' 
    \frac{\bar{\rho}(L,\rp)}{\hat{r}_b(L)} 
    w(\rp) j_\nu(r') Y^*_\nu(\theta,\phi),
    \label{eq:mean}
\ee
the local group contribution is given by
\bm
  \rho_\nu({\rm LG},\rp) = c_\nu \int\!d^3\!r' (\vlg\cdot\hrp)
    \frac{\bar{\rho}(L,\rp)}{\hat{r}_b(L)}  \;\times
\em
\be
    \hspace{2cm}\frac{d}{dr}\left[w(\rp) j_\nu(r')\right]
    Y^*_\nu(\theta,\phi),
    \label{eq:LG}
\ee
and the $\Phi$ and $V$ matrices are defined as
\bm
  \Phi_{\nu\mu} = c_\nu c_\mu \int\!d^3\!r' \bar{\rho}(L,\rp) 
    w(\rp) j_\nu(r') j_\mu(r') \;\times
\em
\be
    \hspace{2cm}Y^*_\nu(\theta,\phi) Y_\mu(\theta,\phi),
    \label{eq:phi}
\ee
and
\bm
  V_{\nu\mu} = \frac{c_\nu c_\mu}{k_\mu^2} \int\!d^3\!r' 
    \frac{\bar{\rho}(L,\rp)}{\hat{r}_b(L)} 
    \frac{\Omega_m(z)^{0.6}}{\Omega_m(0)^{0.6}} D(z) \;\times
\em
\be
    \hspace{1cm}
    \frac{d}{dr'}\left[w(\rp) j_\nu(r')\right] \frac{d}{dr'}j_\mu(r') 
    Y^*_\nu(\theta,\phi) Y_\mu(\theta,\phi).
    \label{eq:V}
\ee

Assuming that the mean observed density field
$\bar{\rho}(L,\rp)=\bar{\rho}(L,r')M(\theta,\phi)$ and the weighting
$w(\rp)=w(r')w(\theta,\phi)$ can be split into angular and radial
components, then the 3D integrals required to calculate the $\Phi$ and
$V$ matrices have the same angular contribution
\be
  W_{\nu\mu} = \int\!d\theta\,d\phi\,w(\theta,\phi)Y^*_\nu(\theta,\phi)
    M(\theta,\phi)Y_\mu(\theta,\phi),
  \label{eq:W}
\ee
where $M(\theta,\phi)$ is the sky mask of the survey. This therefore
only needs to be calculated once. 

The effect of the survey geometry (matrix $\Phi_{\nu\mu}$) is
independent of the luminosity-dependent bias correction: the
$1/\hat{r}_b(L)$ factor in Eq.~\ref{eq:sht} was designed to cancel
the offset in $\delta(L,\rp)$ (Eq.~\ref{eq:biaseddelta}). Note that we
have included the redshift evolution part of $\beta(L_*,z)$ in
Eq.~\ref{eq:V}, and in the calculation performed, so that we fit the
data with $\beta(L_*,0)$. Ignoring this correction gives a measured
$\beta(L_*,z)$ approximately $10\%$ larger than $\beta(L_*,0)$,
because it corresponds to an effective redshift $\sim0.17$.

\subsection{Construction of the covariance matrix}

In this Section we only work with the real space position, and all
overdensities correspond to $L_*$ galaxies. For simplicity, we
therefore define $\delta_\mu\equiv\delta_\mu(L_*,\rp)$. We also define
\be
  \Psi_{\nu\mu} \equiv \sum_\eta 
    \left(\Phi_{\nu\eta}+\beta(L_*,0) V_{\nu\eta}\right)S_{\eta\mu},
  \label{eq:psi}
\ee
so that Eq.~\ref{eq:D1} becomes
\be
  D_\nu = \sum_\mu \Psi_{\nu\mu} \delta_\mu.
    \label{eq:D2}
\ee
The real and imaginary parts of $D_\nu$ are given by
\begin{eqnarray}
  \Real D_\nu &=& \sum_\eta\left(\Real\Psi_{\nu\eta}\Real\delta_\eta
                                -\Imag\Psi_{\nu\eta}\Imag\delta_\eta\right) 
      \label{eq:Dsplit1} \\
  \Imag D_\nu &=& \sum_\eta\left(\Imag\Psi_{\nu\eta}\Real\delta_\eta
                                +\Real\Psi_{\nu\eta}\Imag\delta_\eta\right).
  \label{eq:Dsplit2}
\end{eqnarray}

From Eqns.~\ref{eq:psi} \&~\ref{eq:D2} it can be seen that, for a
single mode, the expected value $\llangle \Real D_\nu \Real
D_\mu\rrangle$ or $\llangle \Imag D_\nu \Imag D_\mu\rrangle$ can be
split into three components dependent on $\beta(L_*,0)^n$ with
$n=0,1,2$.

Given the large number of modes within the linear regime, rather than
estimating the covariances of all modes, we reduce the problem to
considering a number of combinations of the real and imaginary parts
of $D_\nu$. We will discuss how we optimally chose the direction of
the component vectors in the space of all modes in
Section~\ref{sec:compression}. Suppose the revised mode combinations
that we wish to consider are given by
\be
  \hat{D}_a = \sum_\nu E^r_{a\nu}\Real D_\nu + 
	      \sum_\nu E^i_{a\nu}\Imag D_\nu.
  \label{eq:evectors}
\ee
Note that in this Equation $a$ does not represent a triplet of
$\ell$,$m$, \& $n$, but is instead simply an index of the modes
chosen.  Using Equations~\ref{eq:Dsplit1}~\&~\ref{eq:Dsplit2}, we can
decompose into multiples of the real and imaginary components of
$\delta$
\be
  \hat{D}_a = 
    \sum_\eta\left(\Upsilon^r_{a\eta}\Real\delta_\eta
  		  +\Upsilon^i_{a\eta}\Imag\delta_\eta\right),
\ee
where
\begin{eqnarray}
  \Upsilon^r_{a\eta} &=& 
    \sum_\nu\left(E^r_{a\nu}\Real\Psi_{\nu\eta}
		 +E^i_{a\nu}\Imag\Psi_{\nu\eta}\right)
    \label{eq:DefTaur} \\
  \Upsilon^i_{a\eta} &=&
    \sum_\nu\left(E^i_{a\nu}\Real\Psi_{\nu\eta}
		 -E^r_{a\nu}\Imag\Psi_{\nu\eta}\right).
    \label{eq:DefTaui}
\end{eqnarray}
The expected values of $\langle \hat{D}_a \hat{D}_b \rangle$ are
then 
\bm
  \langle \hat{D}_a \hat{D}_b \rangle = 
    \sum_\eta \sum_\gamma \langle
    (\Upsilon^r_{a\eta}\Real\delta_\eta
    +\Upsilon^i_{a\eta}\Imag\delta_\eta) \;\times
\em
\be
    \hspace{2cm}
    (\Upsilon^r_{b\gamma}\Real\delta_\gamma
    +\Upsilon^i_{b\gamma}\Imag\delta_\gamma)\rangle.
    \label{eq:expDD}
\ee
Assuming a standard Gaussian density field, the double sum in
Eq.~\ref{eq:expDD} can be reduced to a single sum using the
following relations
\begin{eqnarray}
  \langle\Real\delta_\nu\Imag\delta_\mu\rangle &=& 0 \\
  \langle\Real\delta_\nu\Real\delta_\mu\rangle &=& 
    \left[\Delta^K_{\nu,\mu}+(-1)^{m_{\nu}}
    \Delta^K_{\nu,-\mu}\right]\frac{P(k_\nu)}{2} \\
  \langle\Imag\delta_\nu\Imag\delta_\mu\rangle &=& 
    \left[\Delta^K_{\nu,\mu}-(-1)^{m_{\nu}}
    \Delta^K_{\nu,-\mu}\right]\frac{P(k_\nu)}{2},
\end{eqnarray}
where the $\Delta^K_{\nu,-\mu}$ terms arise because $\delta_\nu$ obeys
the Hermitian relation $\delta^*_\nu=(-1)^{m_\nu}\delta_{-\nu}$. These
terms are only important for geometries that lack azimuthal symmetry,
such as the 2dFGRS and are less important for the PSCz survey. The
dependence on $P(k)$ follows because the transformation from the
Fourier basis to the Spherical Harmonics basis is unitary and the
amplitude of the complex variable is unchanged. Using
these relations, Eq.~\ref{eq:expDD} reduces to
\bm
  \langle \hat{D}_a \hat{D}_b \rangle = 
    \sum_\eta \frac{P(k_\nu)}{2}\left[
     \Upsilon^r_{a\eta}\Upsilon^r_{b\eta}
    +\Upsilon^i_{a\eta}\Upsilon^i_{b\eta}\right.
\em
\be
    \hspace{1cm}\left.
    +(-1)^{m_{\eta}}\Upsilon^r_{a\eta}\Upsilon^r_{b-\eta}
    -(-1)^{m_{\eta}}\Upsilon^i_{a\eta}\Upsilon^i_{b-\eta}\right].
  \label{eq:cov_geom}
\ee
This equation gives the geometrical component of the covariance matrix
resulting from the mixing of modes caused by the survey geometry and
large-scale redshift-space distortions. Note that, by substituting
Eqns.~\ref{eq:psi}, \ref{eq:DefTaur} \&~\ref{eq:DefTaui} into this
equation we could split the geometric part of the covariance matrix
into 3 components with varying dependence on $\beta(L_*,0)$. This is
actually the case in our implementation of the method so we only have
to calculate these three components once for any value of
$\beta(L_*,0)$.

In addition, there is a shot noise component which can be calculated
by the methods of Peebles (1973). This term enters into the above
formalism because the density field $\rho(L,\sp)$ in Eq.~\ref{eq:sht} is
actually the sum of a series of delta functions, each at the position
of a galaxy. The expected value of $\langle D_\mu D_\nu\rangle$
therefore includes two terms (as in Appendix~A of Feldman, Kaiser \&
Peacock 1994) corresponding to the convolved power and the shot
noise. Allowing $a$ and $b$ to represent either real or imaginary
parts, the expected value of the noise component for each mode, for a
particular galaxy luminosity is
\bm
  \langle aN_\nu\,bN_\mu \rangle = c_\nu c_\mu \int\!d^3\!r 
    \frac{\bar{\rho}(L,\r)}{\hat{r}_b^2(L)} w^2(\r) j_\nu(r) j_\mu(r) r^2 \times
\em
\be
    \hspace{2cm}
    aY^*_\nu(\theta,\phi) bY^*_\mu(\theta,\phi).
\ee
To allow for all galaxy luminosities, we simply integrate over
luminosity as in Percival, Verde \& Peacock (2004).

Allowing for the combinations of modes defined in Eq.~\ref{eq:evectors},
\bm
  \langle \hat{N}_a\,\hat{N}_b \rangle = 
    \sum_\nu\sum_\mu\left(
    E^r_{a\nu}E^r_{b\mu}\langle\Real N_\nu\Real N_\mu\rangle
    \right. 
\em
\bm
    \hspace{1cm}
   +E^i_{a\nu}E^i_{b\mu}\langle\Imag N_\nu\Imag N_\mu\rangle
   +E^r_{a\nu}E^i_{b\mu}\langle\Real N_\nu\Imag N_\mu\rangle
\em
\be
    \hspace{1cm}\left.
   +E^i_{a\nu}E^r_{b\mu}\langle\Imag N_\nu\Real N_\mu\rangle
    \right)
  \label{eq:cov_shot}
\ee

The components of the covariance matrix of the reduced data are
$\langle \hat{D}_a \hat{D}_b \rangle+\langle
\hat{N}_a\,\hat{N}_b \rangle$ as given by
Equations~\ref{eq:cov_geom}~\&~\ref{eq:cov_shot}.

\subsection{Some practical issues}  \label{sec:practical}

\begin{figure}
  \setlength{\epsfxsize}{0.9\columnwidth} 
    \centerline{\epsfbox{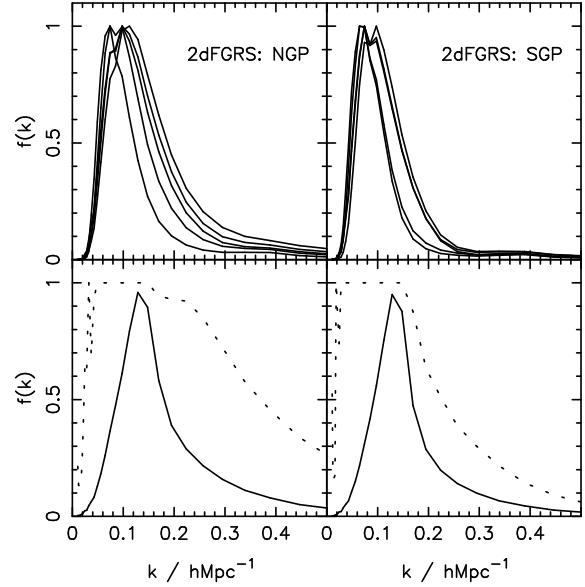}}

  \caption{Normalized contribution to $P(k)$ as a function of $k$ for
  5 example modes for the NGP and SGP (top row). In the lower row we
  present the average (solid line) and maximum (dotted line) of the
  normalized distribution of $k$-contributions, calculated from all
  modes used. \label{fig:win}}
\end{figure}

The calculation of the angular part of the mixing matrices,
$W_{\nu\mu}$ (given by Eq.~\ref{eq:W}) is more CPU intensive than the
calculation of the radial components. Because of this, Tadros et~al.
(1999) utilized Clebsch-Gordan matrices to relate a single transform
of the angular mask to the full transition matrix given by
Eq.~\ref{eq:W}. However at the high $\ell$-values required for the
complex geometry of the 2dFGRS survey it is computationally expensive
to calculate these accurately. Because the integral in Eq.~\ref{eq:W}
can be reduced to a sum over the angular mask, a direct integration
proved stable and computationally faster than the more complicated
Clebsch-Gordan method. At low $\ell$-values both methods agreed to
sufficient precision.

In the large-scale regime $k<0.15\hompc$, and in the regime where the
assumed redshift distribution does not have a significant effect on
the recovered power $k>0.02\hompc$ (see P01), there are $86667$ modes
with $m\ge0$. This statistically complete set includes real and
imaginary modes separately, but only includes modes with $m\ge0$
because $D_\nu$ (Equation~\ref{eq:D1}) obeys the Hermitian relation
and positive and negative $m$-modes are degenerate. The maximum $n$ of
the modes in this set is $33$, and the maximum $\ell$ is $101$.

Obviously we cannot invert a $86667\times86667$ covariance matrix with
each mode as a single element for every model we wish to test, and we
therefore need to reduce the number of modes compared. Another serious
consideration is that many of the modes are nearly degenerate. Because
we can only calculate the components required with finite precision,
nearly degenerate modes often become completely degenerate due to
numerical issues and therefore need to be removed from the analysis:
covariance matrices with negative eigenvalues are unphysical. Removing
degenerate modes is discussed in the context of data compression in the
next Section.

There are two convolutions that we need to perform in order to
determine the covariance matrix, given by
Eqns.~\ref{eq:psi}~\&~\ref{eq:D2}. The number of modes summed when
numerically performing these convolutions is limited by computational
time. The first convolution is given by Equation~\ref{eq:psi} and
results from the small-scale velocity dispersion correction. This
convolution is a simple convolution in $n$ and is relatively narrow in
the linear regime that we consider in this paper. In fact we chose to
convolve over $1\le n\le100$. The second convolution is given by
Equation~\ref{eq:cov_geom}, and is performed for $\ell\le200$. This is
complete for $k<0.29\hompc$, and contains $>4\times10^6$ modes. A
limit in $\ell$ was chosen rather than a limit in $k$ as the CPU time
taken to perform the convolution is dependent on $\ell_{\rm max}$. The
$k$-distribution of contributions to a few of the chosen modes is
presented in Fig.~\ref{fig:win}. Note that although the convolved set
of modes is complete for $k<0.29\hompc$, the fall-off to higher $k$ is
very gentle, and most of the signal beyond this limit will still be
included in the convolution.

\subsection{Data compression}  \label{sec:compression}

As mentioned in Section~\ref{sec:practical}, there are $86667$
Spherical Harmonic modes with $0.02<k<0.15\hompc$, and it is
impractical to use all of these modes in a likelihood
analysis. Consequently, we reject modes for the following reasons
\begin{enumerate}

  \item The 2dFGRS regions considered have a relatively small
  azimuthal angle, so modes that are relatively smooth in this
  direction will be close to degenerate. We therefore set a limit of
  $\ell-|m|>5$ for the modes analysed. This limit effectively
  constrains the number of azimuthal wavelengths in the modes used.

  \item Modes with similar $\ell$-values were found to be closely
  degenerate. Rather than applying a more optimal form of data
  compression, it was decided to simply sample the range of
  $\ell$-values with $\Delta\ell=10$. This spacing was chosen by
  examining the number of small eigenvalues in the three components of
  the covariance matrix as described after Eq.~\ref{eq:cov_geom}.

\end{enumerate}

In addition, we carry out the following steps to remove degenerate
modes in the covariance matrix and to compress the data further. These
steps are performed first in the angular direction (assuming modes
with different $\ell$ and $n$ are independent), then on all of the
remaining modes.

\begin{enumerate}

  \item Even after rejection of near $\ell$-values, nearly degenerate
  combinations of modes remain, which, given the limited numerical
  resolution achievable, could give negative eigenvalues in the
  covariance matrix. Because of this, only modes with eigenvalues in
  the covariance matrix greater than $10^{-5}$ times the maximum
  eigenvalue, well above the round-off error, are retained in the
  three components of the covariance matrix as described after
  Eq.~\ref{eq:cov_geom}. This step is effectively a
  principal-component reduction of the covariance matrix eigenvectors.

  \item Finally, we perform a Karhunen-Lo\`{e}ve decomposition of the
  covariance matrix optimized to constrain $\beta(L_*,0)$. After our
  angular reduction we retain $2155$ \& $2172$ modes for the NGP and
  SGP respectively after this step. Following radial compression we
  are left with $1223$ \& $1785$ modes for the NGP and SGP
  respectively. The number of modes retained for the NGP is smaller
  than for the SGP because the smaller angular coverage means that
  more modes are nearly degenerate.

\end{enumerate}

\subsection{Calculating the likelihood}  \label{sec:like_calc}

Following the hypothesis that $\Real\D_\nu$ and $\Imag\D_\nu$ are
Gaussian random variables, the likelihood function for the variables
of interest can be written 
\bm
  {\cal {L}}[\D|\beta(L_*,0),P(k)] = 
\em
\be
  \hspace{1cm}\frac{1}{(2 \pi)^{N/2} |{\C}|^{1/2}}
    {\rm exp}\left[ - \frac{1}{2} \D^T \C^{-1} \D\right].
    \label{eq:like}
\ee

Matrix inversion is an $N^3$ process, so finding the inverse
covariance matrix can be prohibitively slow in order to test a large
number of models. However, the KL procedure described in
Section~\ref{sec:compression} means that the covariance matrix is
diagonal for a model chosen to be close to the best fit position. To
first order, we might be tempted to assume that the covariance matrix
is diagonal over the range of models to be tested. However, this can
bias the solution depending on the exact form of the matrix. A
compromise is to apply the iterative Newton-Raphson method of
root-finding to matrix inversion (section 2.2.5 of Press et~al. 1992)
starting with the diagonal inverse covariance matrix as the first
estimate. Given an estimate of the inverse covariance matrix $H_0$,
our revised estimate is $H_1=2H_0-H_0CH_0$. Because $H_0$ is diagonal,
the first step of this iterative method only takes of order $N^2$
operations. This trick allows the likelihood to be quickly calculated
for a large number of models, and we use this method in
Section~\ref{sec:results_fixshape}, when we consider a fixed power
spectrum shape.

However, over the larger range of models considered in
Section~\ref{sec:results_varshape}, the covariance matrix changes
significantly, and the estimate $H_1$ is not sufficiently
accurate. Instead, a full matrix inversion is performed for each
model, so mapping the likelihood hypersurface becomes computationally
expensive. A fast method for mapping surfaces which has recently
become fashionable in cosmology is the Markov-chain Monte-Carlo
technique, where an iterative walk is performed in parameter space
seeking local likelihood maxima (e.g. Lewis \& Bridle 2002; Verde et
al. 2003; Tegmark et~al. 2003b). However, we only wish to consider
variation of 4 parameters ($\beta(L_*,z)$, $b(L_*,0)\sigma_8$,
$\Omega_mh$, \&~$\Omega_b/\Omega_m$) in a very simple model described
in Section~\ref{sec:parameters}, so it is easy to map the likelihood
surface using a grid.

\end{document}